\documentclass{jfm}
\usepackage{graphicx}
\usepackage{amsmath}
\usepackage{empheq}
\usepackage{amssymb}
\usepackage{multirow}
\usepackage{color}
\usepackage{subcaption}
\usepackage{float}

\shorttitle{Analytical solution of multiphase unsteady Stokes flow}
\shortauthor{Y. Li, K. Alame and K. Mahesh}
\title{Multiphase unsteady Stokes flow over grooved surfaces: an analytical study}

\author{Yixuan Li\aff{1}, 
 Karim Alame\aff{1}
 \and Krishnan Mahesh\aff{1}
\corresp{\email{kmahesh@umn.edu}}}

\affiliation{\aff{1}Department of Aerospace Engineering and Mechanics, University of Minnesota Twin Cities,
Minneapolis, MN 55455, USA}

\begin{document}

\maketitle

\begin{abstract}
  This paper studies multiphase flow within grooved textures exposed to external unsteadiness. We derive analytical expressions for multiphase unsteady Stokes flow driven by oscillating streamwise/spanwise velocity in the presence of periodic grooves. Good agreement is obtained between the analytical solution and direct numerical simulations performed using the volume of fluid method. The effects of oscillation frequency, Reynolds number, and the multiphase interface location on the transfer function between the input signal external to the groove and output near the interface, are examined. The effective slip length and the shear stress over the grooved plane are described.
\end{abstract}

\section{Introduction}
The potential drag reducing features of superhydrophobic surface (SHS) \citep{quere2008} have renewed interest in the multiphase flow near patterned surfaces. To characterise the SHS, one needs to resolve both its micro-structure and multiphase features, which can be expensive at high Reynolds numbers (\Rey). Therefore, it would be advantageous to predict analytically the flow field near SHS. Pioneered by \citeauthor{philip:1}'s study (\citeyear{philip:1}) of flow over alternating no-slip and no-shear boundary conditions, several aspects of SHS were analytically studied by \cite{lauga2003,Sbragaglia2007,Bazant2008,Crowdy2010,schonecker2013,Luca2018} and many other researchers. 

The typical size of the SHS is often within the viscous sublayer (see \cite{golovin2016} and references therein), therefore the convective terms in the Navier-Stokes equations are negligible as compared to the viscous terms, reducing the governing equations for the fluid to the Stokes equations. Moreover, the inertia terms are comparable to the convective terms, which further simplifies the problem into a quasi-steady Stokes problem or steady problem if the boundary condition is time-dependent \citep{Luchini1991}. Most analytical work on such problems have agreed on this assumption and adopted the steady Stokes equations as the governing equation.

Although the shear-free interface boundary condition has been widely applied \citep{philip:1,philip:2,Ng2009,Sbragaglia2007,Crowdy2010,Schnitzer2017}, there has been an emergence of liquid-infused surface in recent years: \cite{Rosenberg2016,Fu2017,van2017,Gas2017}, which requires a more accurate estimation of the interfacial shear. \cite{belyaev2010,schonecker2013,nizkaya2014gas} have considered the finite shear by determining a finite slip length, placing the interface at the crest of the corrugated surface and treating it as flat. The interface, however, can deform due to capillary effects, as considered in \cite{Sbragaglia2007} and \cite{Crowdy2016,Crowdy2017} using a zero-shear approximation. \cite{Li2017prf} applied an approximate boundary condition to prescribe finite shear and considered interface deformation by placing the interface below the crest of the corrugated surface. Their solution is steady and accurate when the gas layer is thin. In this paper, the penetration effect is included similarly but with both phases solved simultaneously, and the solution being unsteady.

The flow may either be shear-driven \citep{Wang2003,Ng2009,Kamrin2010,schonecker2014,Li2017prf}, or pressure-driven \citep{Sbragaglia2007,belyaev2010,Ng2010}. For shear-driven flow, a shear rate is usually imposed at an infinite height, hence the solution is valid when the channel height is much greater than the geometric features on the surface. Pressure-driven flow can be decomposed into a Poiseuille flow and a perturbation due to the SHS for Stokes flow \citep{Teo2009}, which works well when the shear rate on the multiphase interface is determined heuristically \citep{Teo2009,schonecker2013}. If the second fluid inside the microstructures is included in the description, it could experience the same pressure gradient \citep{Crowdy2017} or be driven by the interfacial stress \citep{Ng2010}, considering that the net cross-sectional mass flux in the second fluid is zero \citep{maynes2007}. The influence of different geometric parameters of the grooves and interface condition is discussed in \cite{Li2017prf}. In this paper, the flow is shear-driven by an oscillating velocity in either streamwise or spanwise direction, with the grooves aligned in the streamwise direction. 

We study the effect of freestream unsteadiness on the \textit{multiphase} flow inside the SHS grooves. We obtain analytical solutions for unsteady Stokes flow over an array of longitudinal grooves with two different fluids of arbitrary density and viscosity. The interfacial stress is finite and obtained as part of the solution. The solution is applicable to both air- and liquid-infused surfaces. \S\ref{sec:longitudinal} and \S\ref{sec:transverse} construct and solve the model problem analytically for longitudinal and transverse flows respectively. In \S\ref{sec:validation}, the analytical solution is validated against VOF simulation results. A parametric study of frequency, Reynolds number, and multiphase interface location is performed in \S\ref{sec:results}. Concluding remarks are made in \S\ref{sec:conclusion}.

\section{Longitudinal flow}
\label{sec:longitudinal}
\begin{figure}
  \center
  \def\svgscale{0.45}
  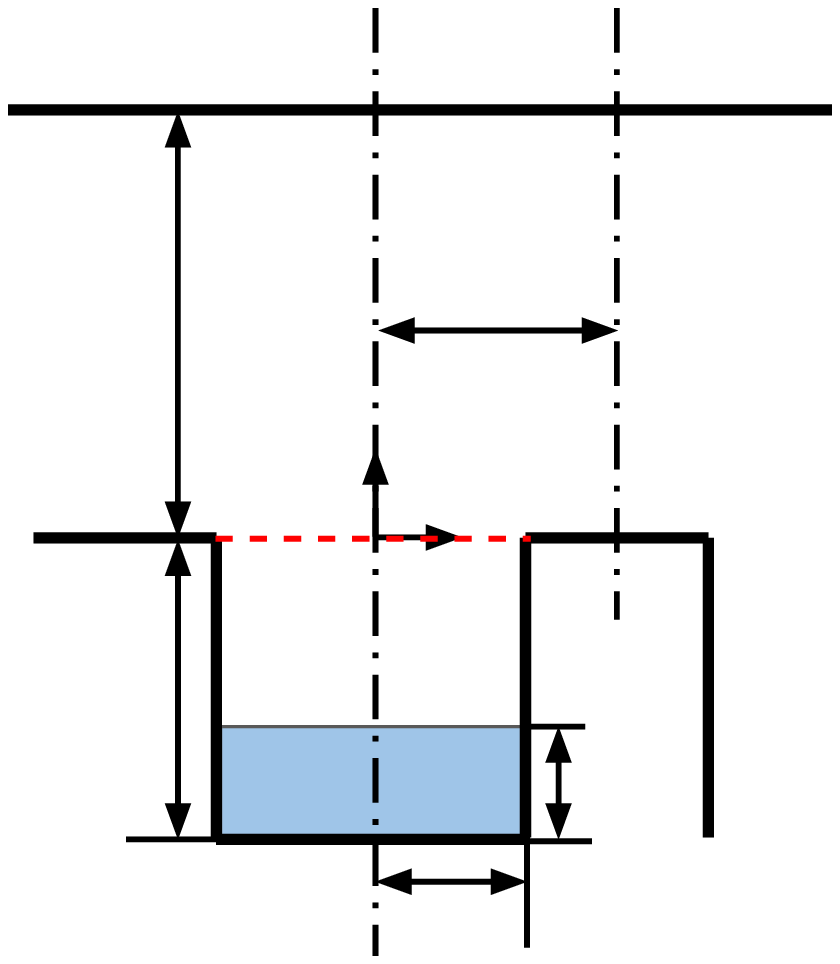
  \caption{\label{fig:def}Schematic of the model problem for longitudinal flow. The red dashed line indicates the boundary between region I and region II. $x$-direction is into the paper.}
\end{figure}

We consider the incompressible and viscous flow over an array of grooves driven by an oscillating velocity. A schematic of the problem is shown in figure~\ref{fig:def}. We introduce the non-dimensional variables as:
\begin{align}
 u &= u^*/U_0,\\
 y &= y^*/L,\\ 
 t &= t^*\omega,
\end{align}
where the variables with the asterisks represent dimensional variables, $u$ is the streamwise velocity, $U_0$ is the velocity magnitude, $L$ is the half period of the groove, and $\omega$ is the frequency of the oscillating velocity. The groove geometry at the lower wall, $y = 0$ is defined by its depth $b$ and half-groove width $a$. The multiphase interface is located at $y=-(b-h)$ and does not change with time. In this section, the velocity is along the direction of the grooves. The oscillating velocity is $u(H,z,t) = \cos(t)$. The groove is periodic in the span and symmetric at its centre.

The dimensionless governing equation is:
\begin{equation}
\frac{\partial u}{\partial t} = \frac{\nu}{L^2\omega}\left(\frac{\partial^2 u}{\partial z^2}+\frac{\partial^2 u}{\partial y^2}\right),
\label{eq:unsteady_stokes}
\end{equation}
where $L^2\omega/\nu$ is the non-dimensional Womersley number. The domain of interest is divided into three rectangular regions. Regions I and III represent the two types of fluid inside the grooves respectively and region II represents the fluid between $y=0$ and the oscillating plane at $y=H$.

For oscillatory flow, we assume the solution is of the form: 
\begin{equation}
u=\Re(\hat{u}\exp(i t)) = \Re(Y(y)Z(z)\exp(i t)).
\label{eq:ufft}
\end{equation}
Substituting \eqref{eq:ufft} into \eqref{eq:unsteady_stokes} we obtain:
\begin{equation}
i\hat{u}=\frac{\nu}{L^2\omega}\left(\frac{\partial^2\hat{u}}{\partial y^2}+\frac{\partial^2\hat{u}}{\partial z^2}\right).
\end{equation}
By separation of variables $\hat{u}(y,z) = Y(y)Z(z)$, we obtain $\frac{Y''}{Y} +\frac{Z''}{Z}=\frac{iL^2\omega}{\nu}$.

We obtain the general solutions as a series of eigensolutions in each region independently and determine the constant coefficients by matching the boundary conditions between the neighbouring regions. The separable solutions are independent of time, therefore the boundary conditions can be written as:
\begin{empheq}[left={ \hat{u}_{\mathrm{II}} =\empheqlbrace}]{align}
\begin{split}
   \hat{u}_{\mathrm{I}},\quad & 0\leqslant z<a, \\
  0,   \quad        & a<z\leqslant 1;
\end{split}
\label{eq:bc1}
\end{empheq}
\begin{equation}
  \frac{\partial \hat{u}_{\mathrm{I}}(0, z)}{\partial y} = \frac{\partial \hat{u}_{\mathrm{II}}(0, z)}{\partial y};\quad 0\leqslant z<a\label{eq:bc2}
\end{equation}
\begin{equation}
  \hat{u}_{\mathrm{I}}(-b+h, z) = \hat{u}_{\mathrm{III}}(-b+h, z);\quad 0\leqslant z<a,\label{eq:bc3}
\end{equation}
\begin{equation}
  \frac{\partial \hat{u}_{\mathrm{I}}(-b+h, z)}{\partial y} = \mu_r\frac{\partial \hat{u}_{\mathrm{III}}(-b+h, z)}{\partial y};\quad 0\leqslant z<a \label{eq:bc4}
\end{equation}
where $\mu_r = \mu_A/\mu_B$ is the viscosity ratio between the external fluid $A$ and the internal fluid $B$ (blue shaded area in figure~\ref{fig:def}), when $\mu_r = 0$, the interface is shear-free. Fluid $A$ can penetrate into the groove. 




\subsection{Region I}
The boundary conditions are no-slip at the wall ($z = a$) and symmetry at the groove centre ($z = 0$). In terms of $\hat{u}(y,z) = Y(y)Z(z)$,
\begin{align}
\frac{\partial\hat{u}_\mathrm{I}(y,0)}{\partial z}&=0 \Rightarrow Z'(0) = 0,\label{eq:41}\\
\hat{u}_\mathrm{I}(y,a)                           &= 0 \Rightarrow Z(a) = 0.\label{eq:51}
\end{align}

From \eqref{eq:41} and \eqref{eq:51}, we get $Z_n(z) = \cos(\alpha_nz)$, where $\alpha_n = (n-\frac{1}{2})\frac{\pi}{a}$. This implies $\frac{Y''}{Y}= \frac{iL^2\omega}{\nu}+\alpha_n^2$. The solution for $Y$ is therefore:
\begin{equation*}
  Y_n(y) = A_{1n}\exp\left(\sqrt{\alpha_n^2+\frac{iL^2\omega}{\nu}}y\right)+A_{2n}\exp\left(-\sqrt{\alpha_n^2+\frac{iL^2\omega}{\nu}}y\right).
\end{equation*}
A separable solution for $\hat{u}_\mathrm{I}$ is:
\begin{align}
  \hat{u}_\mathrm{I}(y, z) &= \sum_{n=1}^{\infty}\cos(\alpha_nz)\left\{A_{1n}\exp\left[\sqrt{\alpha_n^2+\frac{iL^2\omega}{\nu}}y\right]\right.\nonumber\\
  &\left.+A_{2n}\exp\left[-\sqrt{\alpha_n^2+\frac{iL^2\omega}{\nu}}(y+2b-2h)\right]\right\}.
\end{align}
The coefficients $A_{1n}$ and $A_{2n}$ will be obtained by matching adjoining regions below.

\subsection{Region II}
The boundary conditions in region II are symmetric at the centre ($z=0$) and between grooves ($z=1$). Also, at the fixed height $y=H$, the slip velocity oscillates with time.
\begin{align}
\frac{\partial\hat{u}_\mathrm{II}(y,0)}{\partial z}&=0\Rightarrow Z'(0) = 0,\label{eq:13}\\
\frac{\partial\hat{u}_\mathrm{II}(y,1)}{\partial z}&=0\Rightarrow Z'(1) = 0,\label{eq:14}\\
\hat{u}(H,z)&=1.\label{eq:15}
\end{align}

The solution for $Z$ must satisfy equations \eqref{eq:13} and \eqref{eq:14}, i.e.: $Z_n(z) = \cos(\gamma_nz)$, where $\gamma_n = n\pi$.
Hence the solutions for $Y$ must satisfy: $\frac{Y''}{Y}=\frac{iL^2\omega}{\nu}+\gamma_n^2$,
for any integer $n\geqslant 0$. Therefore the eigensolution for $Y$ is:
\begin{equation}
Y_n(y) = B_n\exp\left(\sqrt{\gamma_n^2+\frac{iL^2\omega}{\nu}}y\right)+C_n\exp\left(-\sqrt{\gamma_n^2+\frac{iL^2\omega}{\nu}}y\right).
\end{equation}
Therefore, the general solution in region II is:
\begin{equation}
  \begin{aligned}
    \hat{u}_\mathrm{II}(y, z) = B_0\exp\left[\sqrt{\frac{iL^2\omega}{\nu}}y\right]+C_0\exp\left[-\sqrt{\frac{iL^2\omega}{\nu}}y\right]\\
    +\sum_{n=1}^{\infty}\cos(\gamma_n z)\left[B_n\exp\left(\sqrt{\gamma_n^2+\frac{iL^2\omega}{\nu}}y\right)+ C_n\exp\left(-\sqrt{\gamma_n^2+\frac{iL^2\omega}{\nu}}y\right)\right].
  \end{aligned}
\end{equation}

The inhomogeneous behaviour of the flow in the spanwise direction caused by the groove is represented by $\cos(\gamma_n z)$, which should diminish as the distance to the groove increases; therefore the coefficients $B_n = 0$. Integrating the boundary condition \eqref{eq:15} from $0$ to $1$ yields:
\begin{equation}
  B_0\exp\left[\sqrt{\frac{L^2\omega}{2\nu}}\left(1+i\right)H \right] + C_0\exp\left[-\sqrt{\frac{L^2\omega}{2\nu}}\left(1+i\right)H\right] = 1.
\label{eq:B0C0}
\end{equation}


\subsection{Region III}
The general solution for region III that satisfies the no-slip condition on the walls and symmetry is:
\begin{align}
  \hat{u}_{\mathrm{III}}(y,z) &= \sum_{n = 1}^{\infty}D_n\cos(\alpha_nz)\left[\exp\left(\sqrt{\alpha_n^2+\frac{iL^2\omega}{\nu}}(y+b-h)\right)\right.\nonumber\\
  &\left.-\exp\left(-\sqrt{\alpha_n^2+\frac{iL^2\omega}{\nu}}(y+b+h)\right)\right].
\end{align}

\subsection{Solving for coefficients in the general solutions}
\subsubsection{Between region I and region III}
The boundary at the interface satisfies equations~\eqref{eq:bc3} and \eqref{eq:bc4}.
Multiplying \eqref{eq:bc3} and \eqref{eq:bc4} by $\cos(\alpha_mz)$ and integrating from $0$ to $a$ respectively, yields:
\begin{equation}
  (A_{1m}+A_{2m})\exp\left[-\sqrt{\alpha_m^2+\frac{iL^2\omega}{\nu}}(b-h)\right] = D_m\left[1-\exp\left(-2\sqrt{\alpha_m^2+\frac{iL^2\omega}{\nu}}h\right)\right],
  \label{eq:III}
\end{equation}
and
\begin{equation}
  (A_{1m}-A_{2m})\exp\left[-\sqrt{\alpha_m^2+\frac{iL^2\omega}{\nu}}(b-h)\right] = \mu_rD_m\left[1+\exp\left(-2\sqrt{\alpha_m^2+\frac{iL^2\omega}{\nu}}h\right)\right].
  \label{eq:IIIy}
\end{equation}

\subsubsection{Between region I and region II}
Next, we determine the unknown constants $C_0$, $C_n$ and $D_n$ by applying the boundary condition between the two regions. 
Integrating \eqref{eq:bc1} from 0 to 1,
\begin{align}
B_0+C_0 &= \sum_{n=1}^{\infty}-\frac{(-1)^n}{\alpha_n}\left[A_{1n}+A_{2n}\exp\left(-2\sqrt{\alpha_n^2+\frac{iL^2\omega}{\nu}}(b-h)\right)\right].
    \label{eq:C0}
\end{align}
Multiplying \eqref{eq:bc1} by $\cos(\gamma_mz)$ and integrating from 0 to 1,
\begin{align}
C_m &= 2\sum_{n=1}^{\infty}L_{mn}\left[A_{1n}+A_{2n}\exp\left(-2\sqrt{\alpha_n^2+\frac{iL^2\omega}{\nu}}(b-h)\right)\right],
\label{eq:Cm}
\end{align}
where $L_{mn} = \int_0^a\cos(\alpha_nz)\cos(\gamma_m z)dz$. Multiplying \eqref{eq:bc2} by $\cos(\alpha_mz)$ and integrating from $0$ to $a$,
\begin{align}
&-\frac{(-1)^m}{\alpha_m}\sqrt{\frac{iL^2\omega}{\nu}}(B_0-C_0)-\sum_{n=1}^{\infty}C_nL_{nm}\sqrt{\gamma_n^2+\frac{iL^2\omega}{\nu}}\nonumber\\
&=\frac{a}{2}\sqrt{\alpha_m^2+\frac{iL^2\omega}{\nu}}\left[A_{1m}-A_{2m}\exp\left(-2\sqrt{\alpha_m^2+\frac{iL^2\omega}{\nu}}(b-h)\right)\right].
\label{eq:An}
\end{align}

Finally, $C_m$ is truncated to $M$ terms and $D_n$ is truncated to $N = \mathrm{Int}[aM]$ terms. By solving equations~\eqref{eq:B0C0}, and \eqref{eq:III} to \eqref{eq:An} as a system of equations in MATLAB, we can obtain all the unknown coefficients in the solution. Table~\ref{tab:B0} shows the convergence of $B_0+C_0$ with $M$. $B_0+C_0$ is accurate up to the fourth digit when $M$ is greater than $25$.

\begin{table}
  \centering
\begin{tabular}{l|lllllllll}
$L^2\omega/\nu$\textbackslash M       & 5      & 10     & 15     & 20     & 25     & 30     & 35     & 40     & 45     \\ \hline
0.1875                & 0.0232 & 0.0234 & 0.0234 & 0.0234 & 0.0235 & 0.0235 & 0.0235 & 0.0235 & 0.0235 \\
0.0625                & 0.0366 & 0.0369 & 0.0369 & 0.0369 & 0.0370 & 0.0370 & 0.0370 & 0.0370 & 0.0370
\end{tabular}
\caption{\label{tab:B0}{Convergence of $\Re\left(B_0+C_0\right)$, $a = 0.6$, $b = 1.25$, $h = 0.9b$, $H = 5$, $\mu_r = 0.02$.}}
\end{table}

\section{Transverse flow}
\label{sec:transverse}
\begin{figure}
  \center
  \def\svgscale{0.45}
  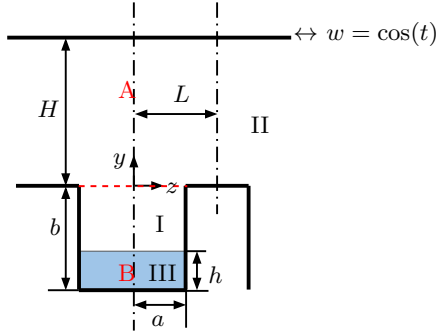
  \caption{\label{fig:def_tran}Schematic of the model problem for transverse flow. The red dashed line indicates the boundary between region I and region II. $x$-direction is into the paper.}
\end{figure}

For transverse flow, the freestream oscillating velocity is in the $z$ direction: $w(H,z,t) = \cos(t)$. A schematic of the problem is shown in figure~\ref{fig:def_tran}. The dimensionless governing equation is:
\begin{equation}
  i\nabla^2\hat\Psi = \frac{\nu}{L^2\omega}\nabla^4\hat\Psi,
\end{equation}
with the streamfunction ($\hat{\Psi}$) to replace $\hat w = \frac{\partial \hat \Psi}{\partial y}$ and $\hat v = -\frac{\partial \hat \Psi}{\partial z}$.
\subsection{Region I}
$\hat{\Psi}$ and its normal derivatives are zero at the walls and symmetric at the groove centre ($z = 0$). A separable solution for $\hat{\Psi}_\mathrm{I}$ is:
\begin{equation}
  \hat{\Psi}_I(y, z) = \sum_{n=1}^{\infty}\cos(\alpha_nz)f_n(y) + \sum_{n=1}^{\infty}C_n\sin(\lambda_n y) \phi_n(\lambda_n;z),
  \label{eq:cI}
\end{equation}
where $\lambda_n = n\pi/(b-h)$, and
\begin{align}
  f_n(y) &= A_{1n}\exp\left(\sqrt{\alpha_n^2+\frac{iL^2\omega}{\nu}}y\right) + A_{2n}\exp\left[-\sqrt{\alpha_n^2+\frac{iL^2\omega}{\nu}}(y+2(b-h))\right]\nonumber \\
         &+ A_{3n}\exp(\alpha_ny) + A_{4n}\exp[-\alpha_n(y+2(b-h))],
\end{align}
\begin{align}
  \phi_n(\lambda_n;z) & = \exp\left(\sqrt{\lambda_n^2+\frac{iL^2\omega}{\nu}}(z-a)\right)+\exp\left(-\sqrt{\lambda_n^2+\frac{iL^2\omega}{\nu}}(z+a)\right)\nonumber \\
  & - \frac{1+\exp\left(-2\sqrt{\lambda_n^2+\frac{iL^2\omega}{\nu}}a\right)}{1+\exp\left(-2\lambda_na\right)}\left[\exp\left(\lambda_n(z-a)\right) + \exp\left(-\lambda_n(z+a)\right)\right].
  \label{eq:phi}
\end{align}

At the side wall
\begin{equation}
\tilde{\Psi}_{Iz}(y,a) = 0.
\end{equation}
Multiplying this by $\sin(\lambda_my)$ and integrating from $-(b-h)$ to $0$ yields:
\begin{align}
  \sum_{n=1}^\infty \alpha_n (-1)^n(A_{1n}K_{1mn}+A_{2n}K_{2mn}+A_{3n}K_{3mn}+A_{4n}K_{4mn}) + \frac{b-h}{2}C_m\phi_m'(\lambda_m;a) = 0,
  \label{eq:Iza}
\end{align}
where
\begin{equation*}
  \begin{aligned}
    K_{1mn} &= \frac{\lambda_m\left[\exp\left(-\sqrt{\alpha_n^2+\frac{iL^2\omega}{\nu}}(b-h)\right)(-1)^m-1\right]}{\alpha_n^2+\lambda_m^2+\frac{iL^2\omega}{\nu}},\\
    K_{2mn} &= \frac{\lambda_m\left[\exp\left(\sqrt{\alpha_n^2+\frac{iL^2\omega}{\nu}}(b-h)\right)(-1)^m-1\right]\exp\left(-2\sqrt{\alpha_n^2+\frac{iL^2\omega}{\nu}}(b-h)\right)}{\alpha_n^2+\lambda_m^2+\frac{iL^2\omega}{\nu}},\\
    K_{3mn} &= \frac{\lambda_m\left[\exp\left(-\alpha_n (b-h)\right)(-1)^m-1\right]}{\alpha_n^2+\lambda_m^2},\\
    K_{4mn} &= \frac{\lambda_m\left[\exp\left(\alpha_n (b-h)\right)(-1)^m-1\right]\exp\left(-2\alpha_n (b-h)\right)}{\alpha_n^2+\lambda_m^2}.
  \end{aligned}
\end{equation*}

The location of the interface is at $y=-(b-h)$ such that $\frac{\partial \hat{\Psi}_I}{\partial z} = 0$, yielding,
\begin{equation}
  \exp\left(-\sqrt{\alpha_n^2+\frac{iL^2\omega}{\nu}}(b-h)\right)(A_{1n}+A_{2n}) + \exp\left(-\alpha_n(b-h)\right)(A_{3n}+A_{4n}) = 0.
  \label{eq:Izh}
\end{equation}
\subsection{Region II}
The boundary conditions in region II are symmetric at the centre ($z=0$) and periodic between grooves ($z=1$). At the fixed height $y=H$, the slip velocity oscillates with time. The boundary condition at $y = H$ is:
\begin{equation}
  \hat{\Psi}_{\mathrm{II}y}(H, z) = 1.\label{eq:cII_top}
\end{equation}

The general solution in region II is:
\begin{equation}
  \begin{aligned}
    \hat{\Psi}_\mathrm{II}(y, z) = E_0\exp\left[\sqrt{\frac{iL^2\omega}{\nu}}y\right]+F_0\exp\left[-\sqrt{\frac{iL^2\omega}{\nu}}y\right] + R_0\\
    +\sum_{n=1}^{\infty}\cos(\gamma_n z)\left[E_n\exp\left(-\gamma_n y\right)+ F_n\exp\left(-\sqrt{\gamma_n^2+\frac{iL^2\omega}{\nu}}y\right)\right].
  \end{aligned}
\end{equation}

Integrating the top boundary condition \eqref{eq:cII_top} from $0$ to $1$ gives:
\begin{equation}
  \sqrt{\frac{iL^2\omega}{\nu}}\left[E_0\exp\left(\sqrt{\frac{iL^2\omega}{\nu}}H \right) - F_0\exp\left(-\sqrt{\frac{iL^2\omega}{\nu}}H\right)\right] = 1.
  \label{eq:IIyH}
\end{equation}

\subsection{Region III}

The general solution for region III that satisfies $\hat{\Psi} = 0$ on the walls and symmetry at the centre is:
\begin{align}
  \hat{\Psi}_{\mathrm{III}}(y,z) &= \sum_{n=1}^{\infty}\cos(\alpha_nz)g_n(y) + \sum_{n=1}^{\infty}\sin[\beta_n (y+b-h)] \phi_n(\beta_n;z),
  \label{eq:cIII}
\end{align}
where $\beta_n = n\pi/h$, 
\begin{align}
  g_n(y) &= B_{1n}\exp\left(\sqrt{\alpha_n^2+\frac{iL^2\omega}{\nu}}(y+b-h)\right) + B_{2n}\exp\left(-\sqrt{\alpha_n^2+\frac{iL^2\omega}{\nu}}(y+b+h)\right)\nonumber \\
         &+ B_{3n}\exp(\alpha_n(y+b-h)) + B_{4n}\exp(-\alpha_n(y+b+h)),
\end{align}
and $\phi_n(\beta_n;z)$ is the same as \eqref{eq:phi} with different eigenvalues $\beta_n$. $\hat{\Psi}_\mathrm{III}$ satisfies:
\begin{align}
  \hat{\Psi}_\mathrm{III}(-b, z) & = 0, \label{eq:cIIIb1}\\
  \hat{\Psi}_{\mathrm{III}y}(-b, z) & = 0, \label{eq:cIIIb2}
\end{align}
at the bottom wall, and 
\begin{equation}
  \hat{\Psi}_{\mathrm{III}z} (y, a) = 0, \label{eq:cIIIb3}
\end{equation}
at the side wall $z = a$.

Equation~\eqref{eq:cIIIb1} yields:
\begin{equation}
  \exp\left(-\sqrt{\alpha_n^2+\frac{iL^2\omega}{\nu}}h\right)(B_{1n}+B_{2n}) + \exp(-\alpha_nh)(B_{3n}+B_{4n}) = 0.
  \label{eq:IIIb}
\end{equation}
Multiplying equation~\eqref{eq:cIIIb2} by $\cos(\alpha_mz)$ and integrating from $0$ to $a$:
\begin{align}
  \frac{a}{2}\left[\sqrt{\alpha_m^2+\frac{iL^2\omega}{\nu}}\exp\left(-\sqrt{\alpha_m^2+\frac{iL^2\omega}{\nu}}h\right)(B_{1m}-B_{2m}) + \right.\nonumber \\
  \left. \alpha_m\exp\left(-\alpha_mh\right)(B_{3m}-B_{4m}) \right]  + \sum_{n=1}^{\infty}D_n\beta_n(-1)^nH_{1mn} = 0,
  \label{eq:IIIyb}
\end{align}
where
\begin{equation}
  H_{1mn} = \frac{\frac{iL^2\omega}{\nu}\alpha_m(-1)^m\left[\exp\left(-2\sqrt{\beta_n^2+\frac{iL^2\omega}{\nu}}a\right)\right]}{(\alpha_m^2+\beta_n^2)(\alpha_m^2+\beta_n^2+\frac{iL^2\omega}{\nu})}.
\end{equation}

Multiplying equation~\eqref{eq:cIIIb3} by $\sin(\beta_m(y+b-h))$ and integrating from $-b$ to $-(b-h)$ gives
\begin{equation}
  \sum_{n=1}^{\infty}\alpha_n(-1)^n(B_{1n}P_{1mn}+B_{2n}P_{2mn}+B_{3n}P_{3mn}+B_{4n}P_{4mn}) + \frac{h}{2}D_m\phi_m'(\beta_m; a) = 0,
  \label{eq:IIIza}
\end{equation}
where
\begin{equation*}
  \begin{aligned}
    P_{1mn} &= \frac{\beta_m\left[\exp\left(-\sqrt{\alpha_n^2+\frac{iL^2\omega}{\nu}}h\right)(-1)^m-1\right]}{\alpha_n^2+\beta_m^2+\frac{iL^2\omega}{\nu}},\\
    P_{2mn} &= \frac{\beta_m\left[\exp\left(\sqrt{\alpha_n^2+\frac{iL^2\omega}{\nu}}h\right)(-1)^m-1\right]\exp\left(-2\sqrt{\alpha_n^2+\frac{iL^2\omega}{\nu}}h\right)}{\alpha_n^2+\beta_m^2+\frac{iL^2\omega}{\nu}},\\
    P_{3mn} &= \frac{\beta_m\left[\exp\left(-\alpha_n h\right)(-1)^m-1\right]}{\alpha_n^2+\beta_m^2},\\
    P_{4mn} &= \frac{\beta_m\left[\exp\left(\alpha_n h\right)(-1)^m-1\right]\exp\left(-2\alpha_n h\right)}{\alpha_n^2+\beta_m^2}.
  \end{aligned}
\end{equation*}
The location of the interface is fixed at $y=-(b-h)$, yielding,
\begin{equation}
  B_{1n}+B_{2n}\exp\left(-2\sqrt{\alpha_n^2+\frac{iL^2\omega}{\nu}}h\right) + B_{3n}+B_{4n}\exp\left(-2\alpha_nh\right) = 0.
  \label{eq:IIIzh}
\end{equation}

\subsection{Solving for coefficients in the general solutions}
\subsubsection{Between region I and region III}
The boundary conditions between region I and region III are:
\begin{equation}
  \hat{\Psi}_{\mathrm{I}y}(-b+h, z) = \hat{\Psi}_{\mathrm{III}y}(-b+h, z),\label{eq:cbc5}
\end{equation}
\begin{equation}
  \hat{\Psi}_{\mathrm{I}yy}(-b+h, z) = \mu_r\hat{\Psi}_{\mathrm{III}yy}(-b+h, z).\label{eq:cbc6}
\end{equation}

Multiplying \eqref{eq:cbc5} and \eqref{eq:cbc6} by $\cos(\alpha_mz)$ and integrating from $0$ to $a$, respectively:
\begin{align}
  \frac{a}{2}f_m'(-(b-h)) + \sum_{n=1}^{\infty}C_n\lambda_nH_{2mn} = \frac{a}{2}g_m'(-(b-h)) + \sum_{n=1}^{\infty}D_n\beta_nH_{1mn}.
  \label{eq:IIIyh}
\end{align}
and
\begin{equation}
  \frac{a}{2}f_n''(-(b-h)) = \mu_r\left[ \frac{a}{2}g_n''(-(b-h)) \right],
  \label{eq:IIIyyh}
\end{equation}
where
\begin{equation}
  H_{2mn} = \frac{\frac{iL^2\omega}{\nu}\alpha_m(-1)^m\left[\exp\left(-2\sqrt{\lambda_n^2+\frac{iL^2\omega}{\nu}}a\right)\right]}{(\alpha_m^2+\lambda_n^2)(\alpha_m^2+\lambda_n^2+\frac{iL^2\omega}{\nu})}.
\end{equation}

\subsubsection{Between region I and region II}

The boundary condition between regions I and II:
\begin{empheq}[left={ \hat{\Psi}_{\mathrm{II}} =\empheqlbrace}]{align}
  \begin{split}
    \hat{\Psi}_{\mathrm{I}},\quad & 0\leqslant z<a, \\ 
    0,   \quad        & a<z\leqslant 1;
  \end{split}
  \label{eq:cbc1}
\end{empheq}

\begin{empheq}[left={ \hat{\Psi}_{\mathrm{II}y} =\empheqlbrace}]{align}
  \begin{split}
    \hat{\Psi}_{\mathrm{I}y},\quad & 0\leqslant z<a, \\ 
    0,   \quad        & a<z\leqslant 1;
  \end{split}
  \label{eq:cbc2}
\end{empheq}

\begin{equation}
  \hat{\Psi}_{\mathrm{II}yy} = \hat{\Psi}_{\mathrm{I}yy};\quad  0\leqslant z\leqslant a, 
  \label{eq:cbc3}
\end{equation}

\begin{equation}
  \hat{\Psi}_{\mathrm{II}yyy} = \hat{\Psi}_{\mathrm{I}yyy};\quad  0\leqslant z\leqslant a. 
  \label{eq:cbc4}
\end{equation}

Integrating equation~\eqref{eq:cbc1} from $0$ to $1$ gives:
\begin{align}
  E_0+F_0+R_0 = \sum_{n=1}^\infty\frac{(-1)^{n+1}}{\alpha_n}\left[A_{1n}+A_{2n}\exp\left(-2\sqrt{\alpha_n^2+\frac{iL^2\omega}{\nu}}(b-h)\right)  \right. \nonumber \\
  \left. + A_{3n} + A_{4n}\exp\left(-2\alpha_n(b-h)\right)\right].
  \label{eq:II0}
\end{align}
Multiplying by $\cos(\gamma_mz)$ and integrating yields:
\begin{align}
  \frac{1}{2}(E_m+F_m) &= \sum_{n=1}^\infty L_{mn}\left[A_{1n}+A_{2n}\exp\left(-2\sqrt{\alpha_n^2+\frac{iL^2\omega}{\nu}}(b-h)\right) \right. \nonumber\\
  &\left. + A_{3n}+A_{4n}\exp(-2\alpha_n(b-h))\right].
  \label{eq:II0cos}
\end{align}
Integrating equation~\eqref{eq:cbc2} from $0$ to $1$ gives:
\begin{align}
  (E_0-F_0)\sqrt{\frac{iL^2\omega}{\nu}} &= \sum_{n=1}^\infty\frac{(-1)^{n+1}}{\alpha_n} \nonumber\\
  &\cdot\left\{\sqrt{\alpha_n^2+\frac{iL^2\omega}{\nu}}\left[A_{1n}-A_{2n}\exp\left(-2\sqrt{\alpha_n^2+\frac{iL^2\omega}{\nu}}(b-h)\right)\right] \right. \nonumber\\
  &\left. + \alpha_n\left[A_{3n}-A_{4n}\exp(-2\alpha_n(b-h))\right]\right\} + \sum_{n=1}^\infty C_n\lambda_nJ_{0n}.
  \label{eq:IIy0}
\end{align}
Multiplying by $\cos(\gamma_m z)$ and integrate equation~\eqref{eq:cbc2} from $0$ to $1$ yields:
\begin{align}
  -\frac{1}{2}\left(\gamma_mE_m+\sqrt{\gamma_m^2+\frac{iL^2\omega}{\nu}}F_m\right) &= \sum_{n=1}^\infty L_{mn}\left\{\sqrt{\alpha_n^2+\frac{iL^2\omega}{\nu}} \right.\nonumber \\
 & \left. \cdot \left[A_{1n}-A_{2n}\exp\left(-2\sqrt{\alpha_n^2+\frac{iL^2\omega}{\nu}}(b-h)\right)\right] \right. \nonumber\\
  &\left. + \alpha_n\left[A_{3n}-A_{4n}\exp(-2\alpha_n(b-h))\right]\right\} + \sum_{n=1}^\infty C_n\lambda_nJ_{mn},
  \label{eq:IIy0cos}
\end{align}
where
\begin{align}
  J_{0n} = \frac{1-\exp\left(-2\sqrt{\lambda_n^2+\frac{iL^2\omega}{\nu}}a\right)}{\sqrt{\lambda_n^2+\frac{iL^2\omega}{\nu}}}
  + \frac{\left[1+\exp\left(-2\sqrt{\lambda_n^2+\frac{iL^2\omega}{\nu}}a\right)\right]\left[1-\exp(2\lambda_na)\right]}{\lambda_n\left[1+\exp(2\lambda_na)\right]},
\end{align}

\begin{align}
  J_{mn} &= -\left\{\frac{iL^2\omega}{\nu}\gamma_m\left[1+\exp(-2\lambda_na)\right]\left(1+\exp\left(-2\sqrt{\lambda_n^2+\frac{iL^2\omega}{\nu}}a\right)\right)\sin(\gamma_ma) \right. \nonumber \\
  & \left. + \left[\lambda_n\left(\lambda_n^2+\gamma_m^2+\frac{iL^2\omega}{\nu}\right)(1-\exp(-2\lambda_na))\left(1+\exp\left(-2\sqrt{\lambda_n^2+\frac{iL^2\omega}{\nu}}a\right)\right) \right. \right.\nonumber \\
  & \left. \left. + \sqrt{\beta_n^2+\frac{iL^2\omega}{\nu}}\left(\lambda_n^2+\gamma_m^2\right)(1+\exp\left(-2\lambda_na\right))\left(1-\exp\left(-2\sqrt{\lambda_n^2+\frac{iL^2\omega}{\nu}}a\right)\right)\right]\right. \nonumber \\
  & \left. \cos(\gamma_ma)\right\}/\left\{\left[1+\exp(-2\lambda_na)\right](\lambda_n^2+\gamma_m^2)\left(\lambda_n^2+\gamma_m^2+\frac{iL^2\omega}{\nu}\right)\right\}.
\end{align}
Multiplying equation~\eqref{eq:cbc3} and equation~\eqref{eq:cbc4} by $\cos(\alpha_mz)$ and integrating from $0$ to $a$ yields:
\begin{align}
  &\frac{a}{2}\left\{\left(\alpha_m^2+\frac{iL^2\omega}{\nu}\right)\left[A_{1m}+A_{2m}\exp\left(-2\sqrt{\alpha_m^2+\frac{iL^2\omega}{\nu}}(b-h)\right)\right] \right. \nonumber\\
  &\left. + \alpha_m^2 \left[A_{3m}+A_{4m}\exp\left(-2\alpha_m(b-h)\right)\right]\right\}  \nonumber \\
  &= \frac{iL^2\omega}{\nu}(E_0+F_0)\frac{(-1)^{m+1}}{\alpha_m} + \sum_{n=1}^\infty L_{nm}\left[\gamma_n^2E_n+\left(\gamma_n^2+\frac{iL^2\omega}{\nu}\right)F_n\right],
  \label{eq:IIyy0}
\end{align}
and
\begin{align}
  &\frac{a}{2}\left\{\left(\sqrt{\alpha_m^2+\frac{iL^2\omega}{\nu}}\right)^3\left[A_{1m}-A_{2m}\exp\left(-2\sqrt{\alpha_m^2+\frac{iL^2\omega}{\nu}}(b-h)\right)\right] \right. \nonumber \\
  &\left. +\alpha_m^3\left(A_{3m}-A_{4m}\exp\left(-2\alpha_m(b-h)\right)\right)\right\} \nonumber \\
  &= \left(\sqrt{\frac{iL^2\omega}{\nu}}\right)^3(E_0-F_0)\frac{(-1)^{m+1}}{\alpha_m} - \sum_{n=1}^\infty L_{nm}\left[\gamma_n^3E_n+\left(\sqrt{\gamma_n^2+\frac{iL^2\omega}{\nu}}\right)^3F_n \right],
  \label{eq:IIyyy0}
\end{align}
respectively.

Finally, $E_m$ and $F_m$ are truncated to $M$ terms, $A_n$ and $B_n$ are truncated to $N = \mathrm{Int}[aM]$ terms, $C_n$ and $D_n$ are truncated to $K = \mathrm{Int}[bM]$ terms and solved from a system of equations constructed with equations~\eqref{eq:Iza}, \eqref{eq:Izh}, \eqref{eq:IIyH}, \eqref{eq:IIIb}, \eqref{eq:IIIyb}, \eqref{eq:IIIza}, \eqref{eq:IIIzh}, \eqref{eq:IIIyh}, \eqref{eq:IIIyyh}, \eqref{eq:II0} to \eqref{eq:IIy0cos}, \eqref{eq:IIyy0}, \eqref{eq:IIyyy0} in MATLAB. Table~\ref{tab:EF0} shows the convergence of averaged velocity magnitude $\Re\left(\sqrt{\frac{iL^2\omega}{\nu}}(E_0-F_0)\right)$ at $y = 0$ with $M$. $\Re\left(\sqrt{\frac{iL^2\omega}{\nu}}(E_0-F_0)\right)$ is accurate up to the fourth digit when $M$ is greater than $30$.

\begin{table}
  \centering
\begin{tabular}{l|lllllllll}
$L^2\omega/\nu$\textbackslash M       & 10      & 20     & 30     & 40     & 50     & 60     & 70     \\ \hline
0.1875                & 0.0069 & 0.0068 & 0.0068 & 0.0068 & 0.0067 & 0.0067 & 0.0067 \\
0.0625                & 0.0102 & 0.0101 & 0.0101 & 0.0100 & 0.0100 & 0.0100 & 0.0100 
\end{tabular}
\caption{\label{tab:EF0}{Convergence of $\Re\left(\sqrt{\frac{iL^2\omega}{\nu}}(E_0-F_0)\right)$, $a = 0.6$, $b = 1.25$, $h = 0.9b$, $H = 5$, $\mu_r = 0.02$.}}
\end{table}

\section{Validation}\label{sec:validation}
The solution is first validated against DNS data using the numerical methodology of \cite{mahesh2004}, implemented with VOF given in \cite{Li2017prf}. Then the lower bound of the solution, representing a steady flow result is validated using the analytical solution in \cite{philip:2}, \cite{schonecker2014}, and the simulation result in \cite{Fu2017}.

\subsection{Validation of the multiphase unsteady solution}
The multiphase unsteady solution is validated by DNS with VOF. The results are expressed in the form of the transfer equation defined as:

\begin{empheq}[left={ \mathcal H\left(\frac{L^2\omega}{\nu}\right) = \sqrt{\frac{\Phi(\frac{iL^2\omega}{\nu})_\mathrm{output}}{\Phi(\frac{iL^2\omega}{\nu})_\mathrm{input}}} = \empheqlbrace}]{align}
\begin{split}
  &\sqrt{\hat{u}_\mathrm{I}\hat{u}_\mathrm{I}^*}, \quad \text{longitudinal}, \\
  &\sqrt{\hat{w}_\mathrm{I}\hat{w}_\mathrm{I}^*}, \quad \text{transverse},
\end{split}
\end{empheq}
where the input is considered as the energy of the forcing, and the output is the energy at $y = 0$. 

\begin{figure}
  \centering
  \begin{subfigure}{0.45\textwidth}
    \includegraphics[scale=0.45]{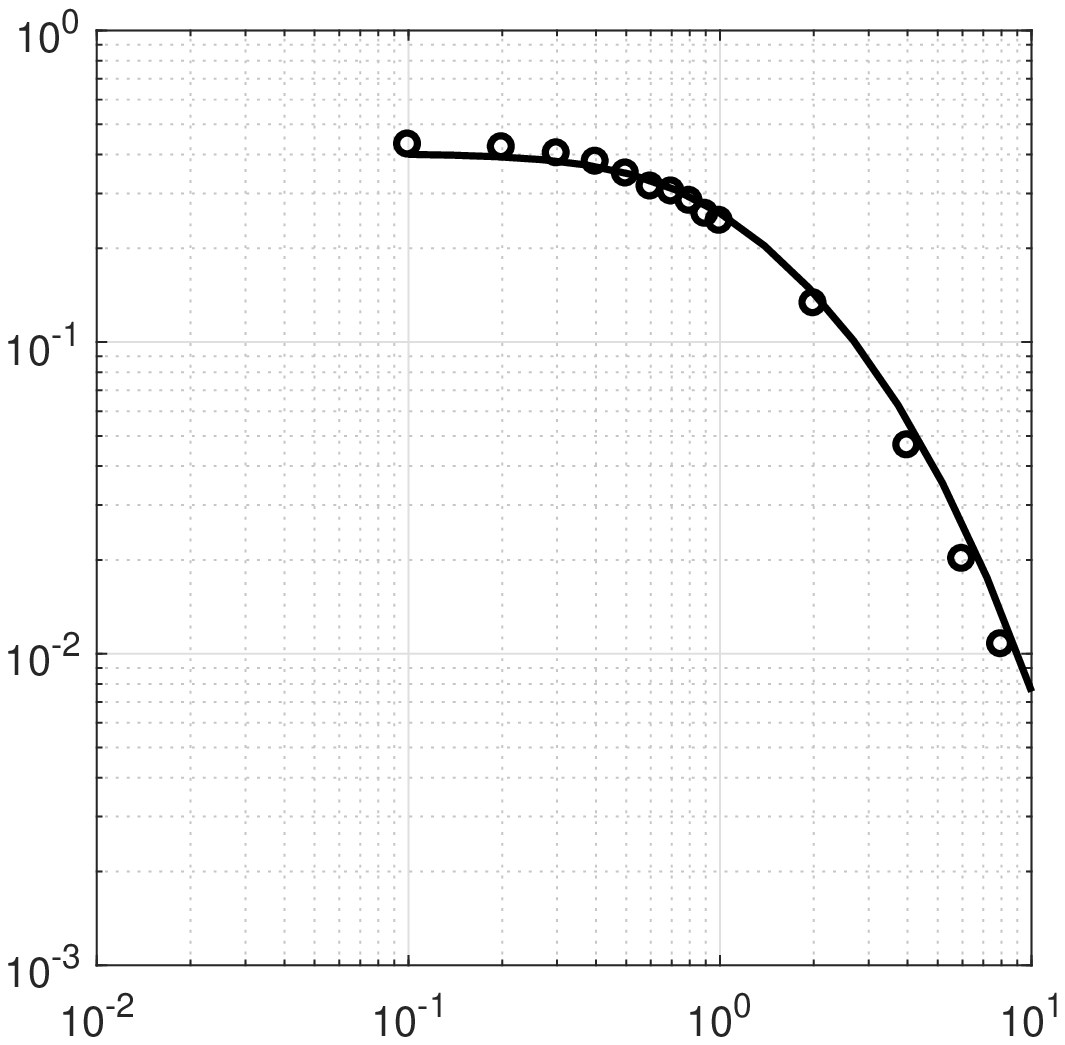}
    \put(-180,70){\rotatebox{90}{$\mathcal H (L^2\omega/\nu)$}}
    \put(-110,-0.5){$L^2\omega/\nu$}
    \caption{Longitudinal flow}
  \end{subfigure}
  \begin{subfigure}{0.45\textwidth}
    \includegraphics[scale=0.45]{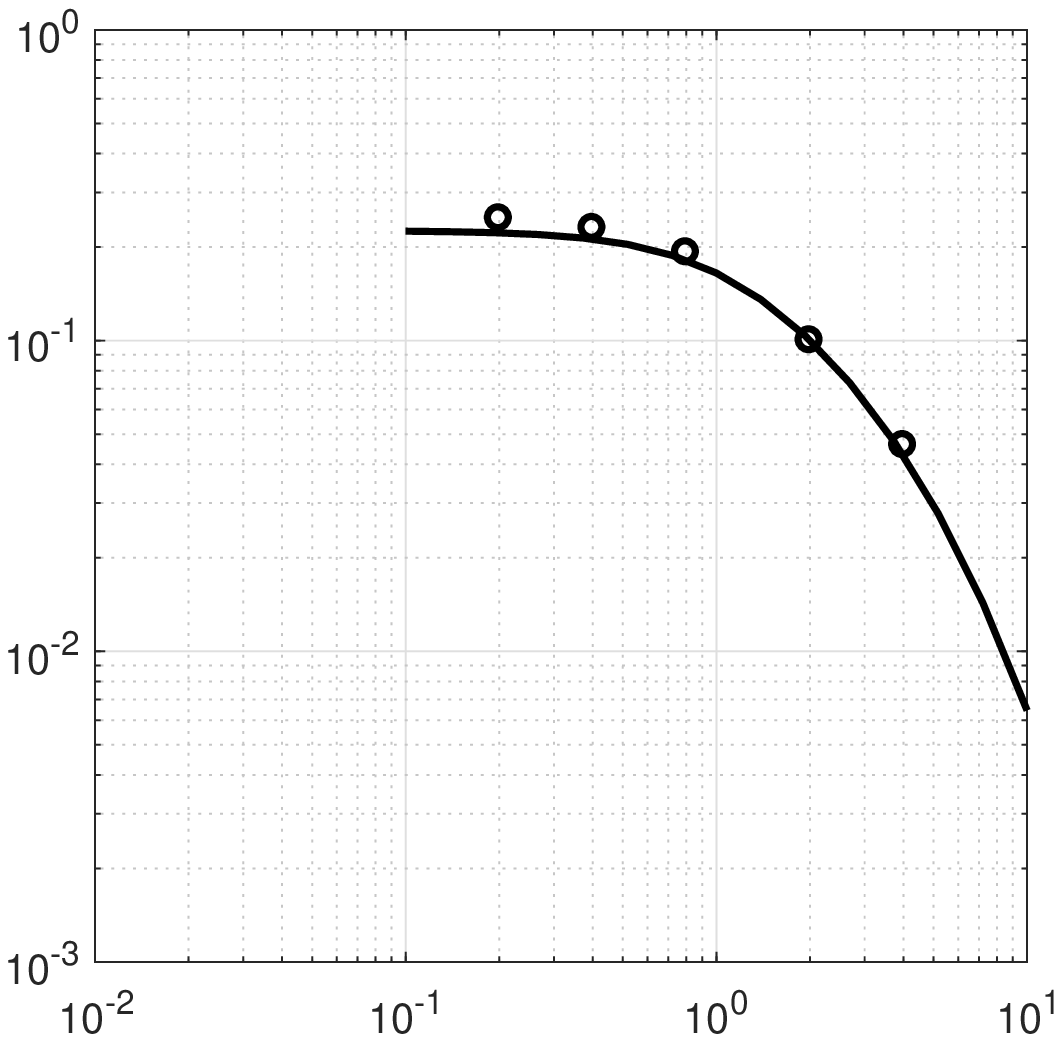}
    \put(-180,70){\rotatebox{90}{$\mathcal H (L^2\omega/\nu)$}}
    \put(-110,-0.5){$L^2\omega/\nu$}
    \caption{Transverse flow}
  \end{subfigure}
  \begin{subfigure}{0.45\textwidth}
    \includegraphics[scale=0.45]{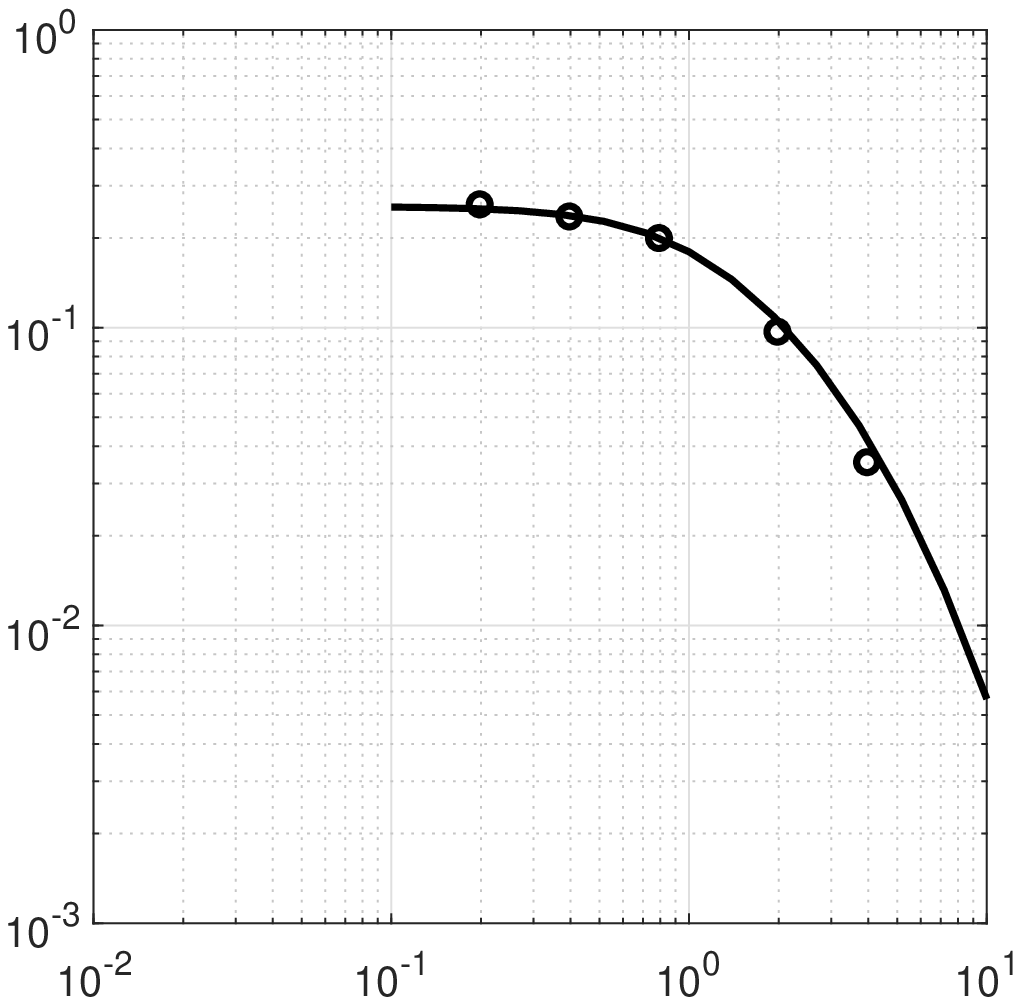}
    \put(-180,70){\rotatebox{90}{$\mathcal H (L^2\omega/\nu)$}}
    \put(-110,-0.5){$L^2\omega/\nu$}
    \caption{Longitudinal flow}
  \end{subfigure}
  \begin{subfigure}{0.45\textwidth}
    \includegraphics[scale=0.48]{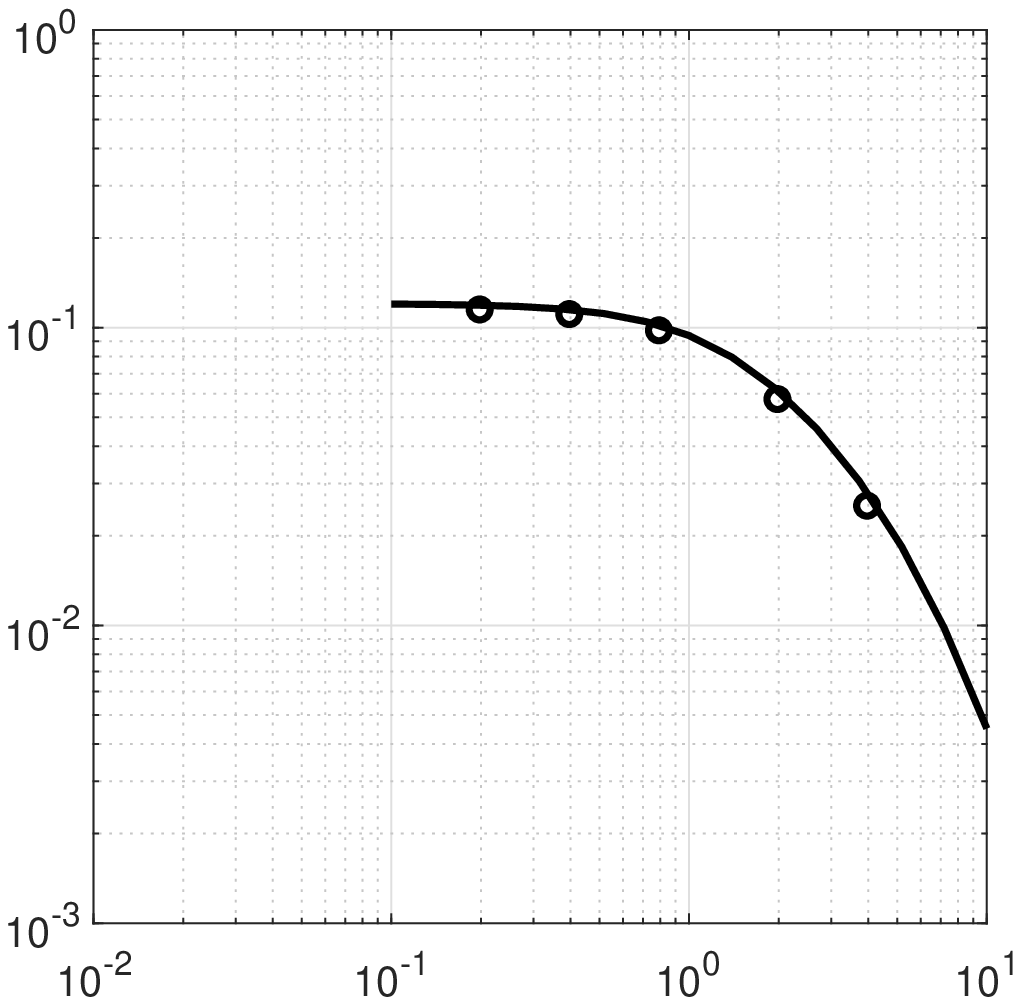}
    \put(-180,70){\rotatebox{90}{$\mathcal H (L^2\omega/\nu)$}}
    \put(-110,-0.5){$L^2\omega/\nu$}
    \caption{Transverse flow}
  \end{subfigure}
    \caption{Comparison of the transfer function results from VOF (symbols) with analytical solutions (solid line). $L=1$, $\nu=1$, $H=2.5$, $a=0.875$, $b = h = 1.75$, and (a, b) $\mu_r = 0.02$, (c, d) $\mu_r = 0.37$.}
  \label{fig:VOF_valid}
\end{figure}

The transfer function shows good agreement between VOF simulation with flat interface ($\sigma\to\infty$) and the analytical solution in figure~\ref{fig:VOF_valid}. In all figures, $L^2\omega/\nu$ varies from $0.1$ to $1.0$, $L=1$, $\nu=1$, $H=2.5$, $h/b = 1.0$, but $\mu_r = 0.02$ in figure~\ref{fig:VOF_valid}(a, b), and $\mu_r = 0.37$ in figure~\ref{fig:VOF_valid}(c, d), respectively.

\subsection{Validation of the multiphase steady solution}

\begin{figure}
  \centering
  \begin{subfigure}{0.45\textwidth}
    \centering
    \includegraphics[scale=0.26]{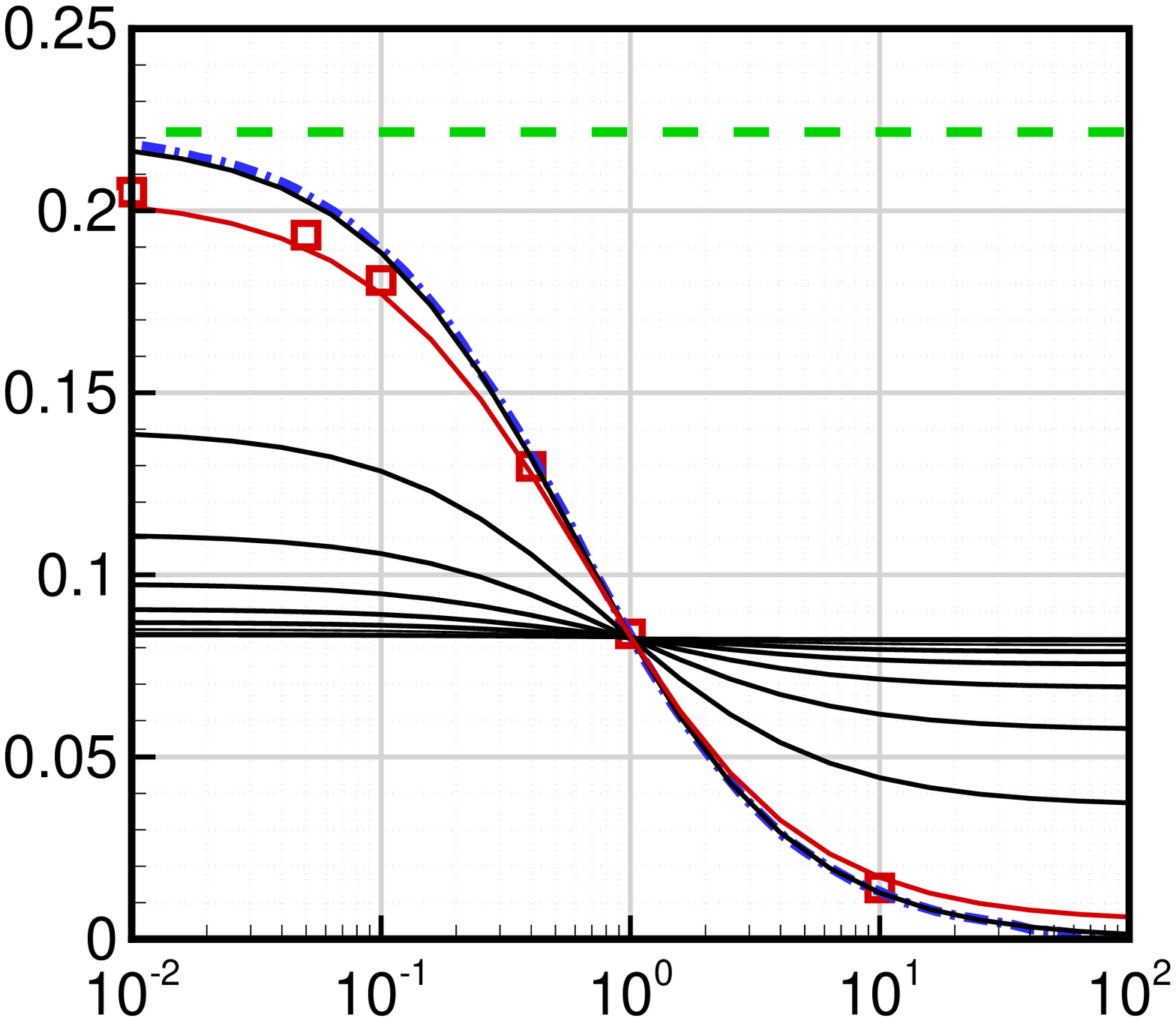}
    \put(-160,60){\rotatebox{90}{$b_\mathrm{eff\parallel}$}}
    \put(-80,0){$\mu_r$}
    \put(-110,110){\vector(-1,-4){20}}
    \caption{Longitudinal flow}
  \end{subfigure}
  \begin{subfigure}{0.45\textwidth}
    \centering
    \includegraphics[scale=0.26]{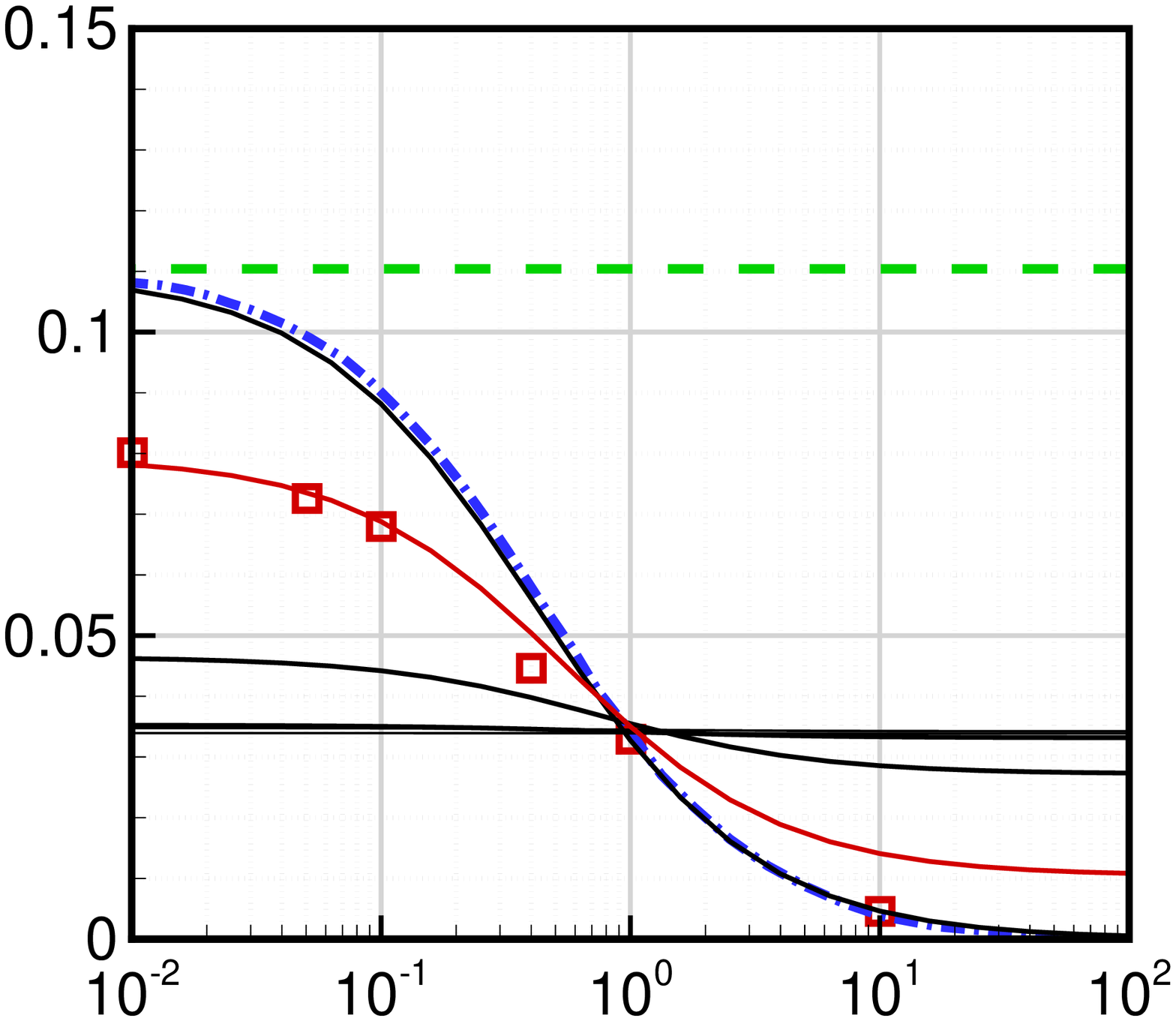}
    \put(-160,60){\rotatebox{90}{$b_\mathrm{eff\perp}$}}
    \put(-80,0){$\mu_r$}
    \put(-110,110){\vector(-1,-4){20}}
    \caption{Transverse flow}
  \end{subfigure}
  \caption{Effective slip length (a) $b_\mathrm{eff\parallel}$ and (b) $b_\mathrm{eff\perp}$ as a function of viscosity ratio $\mu_r$ and the location of the interface $h$. Green dashed line: \cite{philip:2}, blue dashed-dotted line: \cite{schonecker2014}, open square symbols: \cite{Fu2017}, black solid lines: steady solutions derived in this paper. Arrow indicates the decreasing of the interface height with respect to the groove depth $h/b$ ranging from (a) 100\% to 0\% by a step of 10\%; (b) 100\% to 50\% by a step of 10\%. The red line in each plot is (a) $h/b = 99\%$; (b) $h/b=97.5\%$. Data obtained from \cite{Fu2017} is re-normalised in the same manner as this paper.}
  \label{fig:mur_beff}
\end{figure}

\cite{schonecker2014} considered the flow over an array of grooves being filled with a secondary immiscible fluid driven by a constant shear stress $\tau_\infty$ as $y\to\infty$ and derived analytical expressions for effective slip length, with respect to the viscosity ratio of the two fluids. Here, $b_\mathrm{eff} \equiv \overline{u}_\mathrm{II}/ \overline{\partial u_\mathrm{II}/\partial y}$, where the overline `$\overline{()}$' represents the averaged quantity across time and span, with both values evaluated at $y = 0$. This solution can serve as a lower bound of the current unsteady solution where $\omega\to 0$ and $y\to \infty$. This asymptotic trend is confirmed by first comparing two sets of analytical solutions with $L^2\omega/\nu = 10^{-4}$, $H = 5$; and $L^2\omega/\nu = 10^{-5}$, $H = 10$, with the same $a=0.5$, $b=1.0$, $h=b$, and $\mu_r=0.02$. For the two sets of parameters $b_\mathrm{eff\parallel} = 0.2101$, agreeing up to the $10th$ digit. $b_\mathrm{eff\perp} = 0.1032$, agreeing up to the $10th$ digit as well. Therefore, $L^2\omega/\nu = 10^{-4}$, and $H = 5$ are used to compare to the steady flow solutions.

When a small amount of penetration is considered, the data shows good comparison to the simulation data of \cite{Fu2017} at $\mu_r<1$.

\section{Results and discussion}
\label{sec:results}
\subsection{Instantaneous flow field}
\label{sec:inst_flow_field}
\begin{figure}
  \centering
  \includegraphics[scale=0.57]{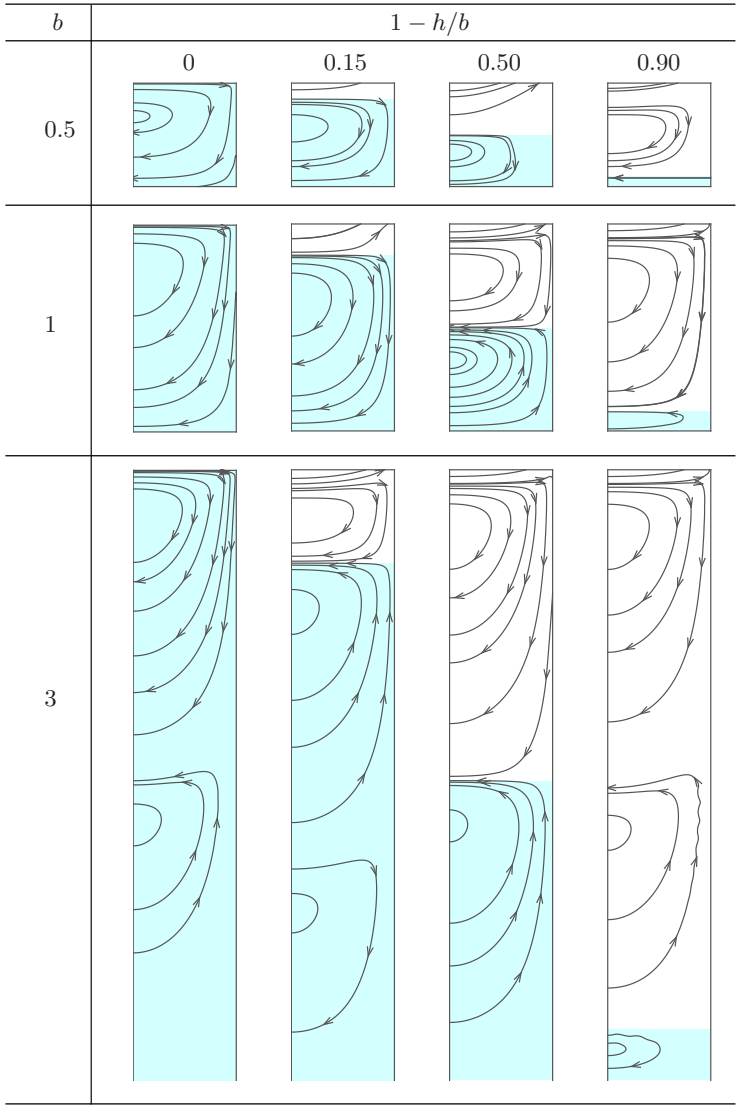}
  \put(-130,403){$1-h/b$}
  \put(-208,388){$0$}
  \put(-155,388){$0.15$}
  \put(-97,388){$0.50$}
  \put(-37,388){$0.90$}
  \put(-257,403){$b$}
  \put(-260,363){$0.5$}
  \put(-260,290){$1$}
  \put(-260,150){$3$}
  \caption{Instantaneous streamlines inside the grooves of different depths $b$ with different penetration rates $1-h/b$ at $t=\frac{2k\pi}{\omega}$, $k$ is an integer. The blue area represents fluid $B$ of $\mu_r=0.02$. The coverage ratio of the groove is $\phi = 0.5$. $L^2\omega/\nu = 0.3428$.}
  \label{fig:inst_plots_table}
\end{figure}

When the grooves are in parallel to the flow direction, the streamwise velocity changes monotonically along the wall normal direction. Its flow pattern has been shown in multiple literature (to name a few: \cite{schonecker2013, Li2017prf}). When the grooves are transverse to the flow direction, the flow field inside the grooves behaves similarly to a lid-driven cavity flow. Figure~\ref{fig:inst_plots_table} shows the instantaneous streamlines inside the grooves of different depths $b$ with different penetration rates $1-h/b$. The external oscillatory boundary condition is all applied at the same $H = 5.40$. The penetration rate represents the area fraction of fluid $A$ inside the groove to the area of the groove $(w(b-h)/wb)$. On one hand, as the groove becomes deeper, multiple vortices form inside, which is similar to the eddy structure in lid-driven cavities with increasing height \citep{Shankar2007}. On the other hand, fluid $B$ in the blue shaded area generates vortices separately from fluid $A$ (white area). Multiple vortices form within fluid $B$ when its depths gets deep as well.

\subsection{Parametric study}
\begin{table}
  \centering
  \begin{tabular}{lcccc}
    $H$    & Penetration rate         & $\frac{L^2\omega}{\nu}$  & $\mu_r$               & $Re_\tau$  \\
    $5.40$ & $0,\ 0.15,\ 0.5,\ 0.9$ & $0.04-4.23$              & $0.02,\ 0.37,\ 30.00$ & $180$     \\
    $2.43$ & $0,\ 0.15,\ 0.5,\ 0.9$ & $0.04-4.23$              & $0.02,\ 0.37,\ 30.00$ & $400$     \\
    $1.65$ & $0,\ 0.15,\ 0.5,\ 0.9$ & $0.04-4.23$              & $0.02,\ 0.37,\ 30.00$ & $590$     
  \end{tabular}
  \caption{Location of the forcing $H$, the penetration rate $1-h/b$, the relevant non-dimensional number $\frac{L^2\omega}{\nu}$, the viscosity ratio $\mu_r$ for all cases solved from analytical solution, and the applicable $Re_\tau$. For all cases, $L=18/3500$ is considered.}
  \label{tab:parametric_study}
\end{table}

In this section, the effect of frequency, domain height, penetration rate, and viscosity ratio are studied using the analytic solution. All the following cases have a coverage ratio of $a=0.875$ otherwise stated. The parameters for each case are listed in table~\ref{tab:parametric_study}. The frequency range is selected to be equivalent to the nondimensional frequency range inside a turbulent channel flow at the given $Re_\tau$ listed on the last column. The location of the forcing $H$ is equivalent to $y^+=5$ for the $Re_\tau$ given. The viscosity ratios are $0.02,\ 0.37,\ 30.00$ to represent air, heptane, and Dupont Krytox to water, respectively \citep{Rosenberg2016}. 

\begin{figure}
  \begin{subfigure}{\textwidth}
    \centering
    \includegraphics[scale=0.39]{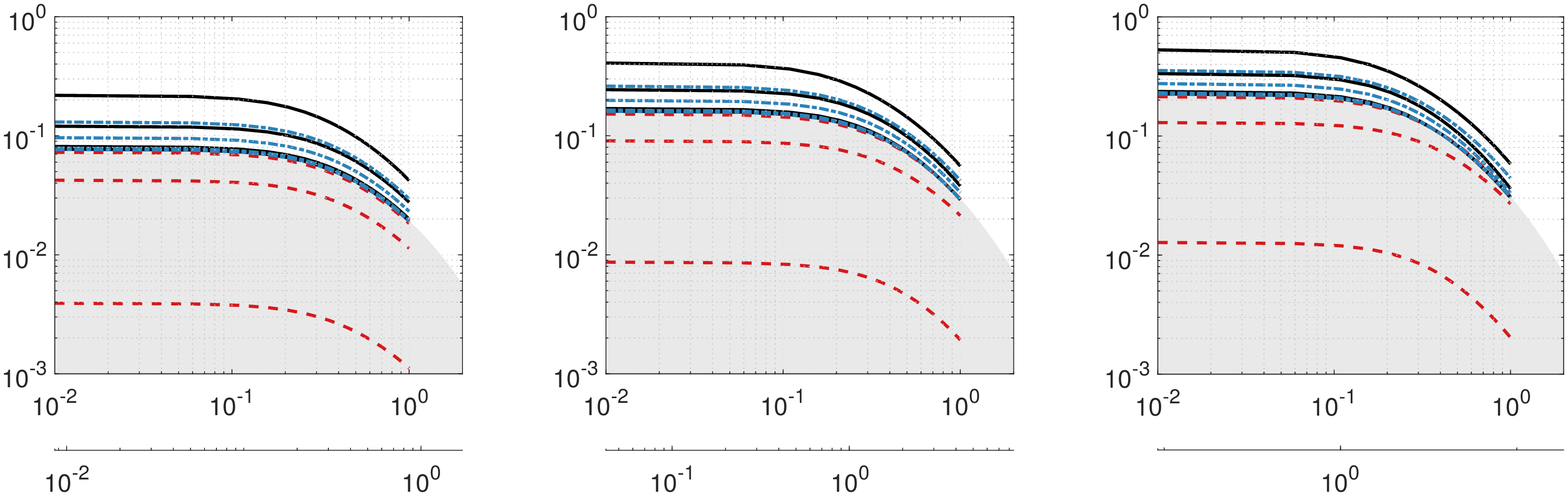}
    \put(-390,50){\rotatebox{90}{$\mathcal H(L^2\omega/\nu)_{uu}$}}
    \put(-320,-10){$L^2\omega/\nu$}
    \put(-190,-10){$L^2\omega/\nu$}
    \put(-60,-10){$L^2\omega/\nu$}
    \put(-320,12){$\omega\nu/u_\tau^2$}
    \put(-190,12){$\omega\nu/u_\tau^2$}
    \put(-60,12){$\omega\nu/u_\tau^2$}
    \put(-350,105){\vector(0,-1){10}}
    \put(-350,65){\vector(0,1){10}}
    \put(-220,110){\vector(0,-1){10}}
    \put(-220,70){\vector(0,1){10}}
    \put(-90,113){\vector(0,-1){10}}
    \put(-90,73){\vector(0,1){10}}
    \put(-280,103){(a)}
    \put(-150,103){(b)}
    \put(-18,103){(c)}
  \end{subfigure}
  \begin{subfigure}{\textwidth}
    \centering
    \includegraphics[scale=0.39]{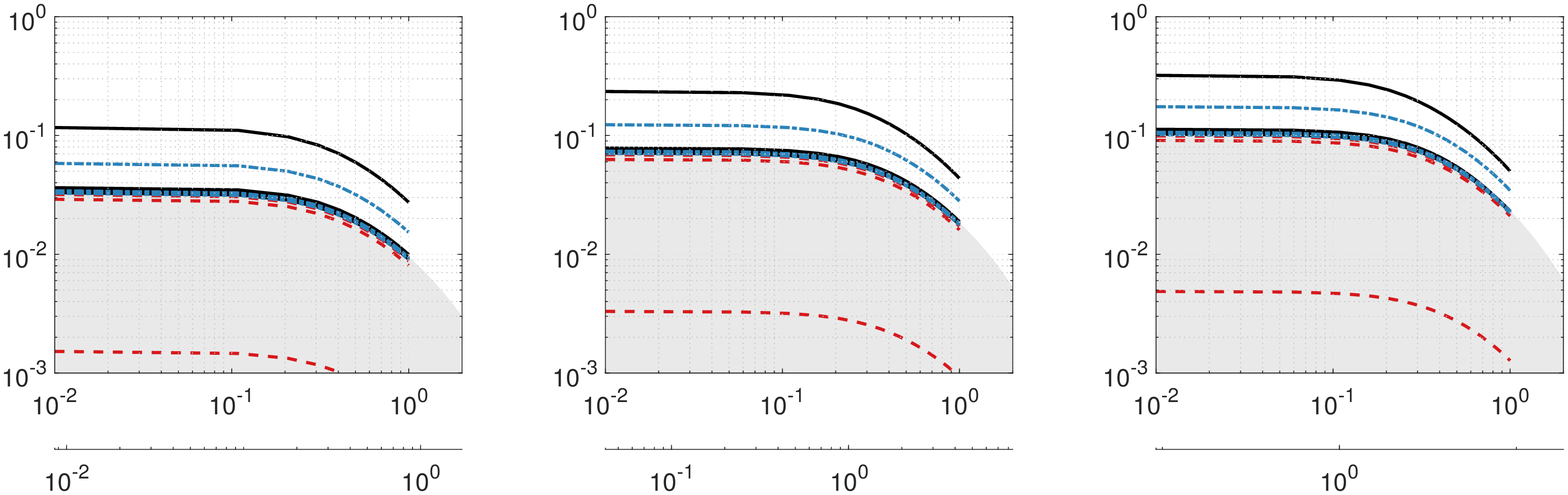}
    \put(-390,50){\rotatebox{90}{$\mathcal H(L^2\omega/\nu)_{ww}$}}
    \put(-320,-10){$L^2\omega/\nu$}
    \put(-190,-10){$L^2\omega/\nu$}
    \put(-60,-10){$L^2\omega/\nu$}
    \put(-320,12){$\omega\nu/u_\tau^2$}
    \put(-190,12){$\omega\nu/u_\tau^2$}
    \put(-60,12){$\omega\nu/u_\tau^2$}
    \put(-350,98){\vector(0,-1){10}}
    \put(-350,58){\vector(0,1){10}}
    \put(-220,105){\vector(0,-1){10}}
    \put(-220,65){\vector(0,1){10}}
    \put(-90,110){\vector(0,-1){10}}
    \put(-90,70){\vector(0,1){10}}
    \put(-280,103){(d)}
    \put(-150,103){(e)}
    \put(-18,103){(f)}
  \end{subfigure}
  \caption{Transfer function $\mathcal H(L^2\omega/\nu)$ in longitudinal flow (top row) and transverse flow (bottom row) with increasing frequencies of the forcing normalized by viscous units $u_\tau^2/\nu$ (top axis) or Womersley number $L^2\omega/\nu$ (bottom axis) with increasing representative (a) and (d) $Re_\tau=180$, (b) and (e) $Re_\tau=400$, (c) and (f) $Re_\tau=590$. Here, $\mu_r = 0.02,\ 0.37,\ 30.00$ are denoted by black solid lines, blue dashed-dotted lines, and red dashed lines respectively. The shaded area represents $\mathcal H(L^2\omega/\nu, \mu_r\geq 1)$. The arrows indicate increasing penetration rate in the area of $\mu_r\geq 1$ or $\mu_r<1$ respectively.}
  \label{fig:transfer_function}
\end{figure}
The transfer function of the cases from table~\ref{tab:parametric_study} are shown in figure~\ref{fig:transfer_function}(a-c). The transfer function $\mathcal H\left(\frac{iL^2\omega}{\nu}\right)$ is defined as the ratio between the energy at $y=0$ and the energy of the forcing at $y=H$:
\begin{equation}
  \mathcal H\left(\frac{L^2\omega}{\nu}\right) = \sqrt{\frac{\Phi(\frac{iL^2\omega}{\nu})|_{y=0}}{\Phi(\frac{iL^2\omega}{\nu})|_{y=H}}}.
\end{equation}

Overall, as the oscillation frequency increases, the transfer function decreases. As the representative Reynolds number increases, the transfer function increases. However, note that as Reynolds number increases, the fluctuations of the forcing would also be closer to the surface. The shaded areas correspond to the condition $\mathcal H(L^2\omega/\nu, \mu_r\geq 1)$ to distinguish between high viscosity fluid and low viscosity fluid inside the groove. Note that the trend of $\mathcal H$ with penetration rate is different between shaded and non-shaded regions. When $\mu_r>1$, as the penetration rate increases, $\mathcal H$ increases, and the penetration of the external fluid improves the transfer of forcing energy. Conversely, when $\mu_r<1$, as the penetration rate increases, $\mathcal H$ decreases, and the penetration of the external fluid suppresses the transfer of the forcing energy. Neither of these two conditions, however, cross the border of the $\mu_r = 1$ curve. Also, air performs the best over the other two fluids, which is intuitively reasonable, considering that the interface does not break or drain based on the assumption made in this paper. The transfer function for transverse flow (figure~\ref{fig:transfer_function} d-f) behaves the same way as in longitudinal flow.

The root-mean-square (rms) of the unsteady analytical effective slip length is determined by:
\begin{equation}
  b_\mathrm{eff\parallel} = \frac{\int_{0}^{1}u(0, z)_\mathrm{rms}dz}{\int_{0}^{1}\frac{\partial u(\infty, z)}{\partial y}_\mathrm{rms}dz} = \frac{\Re\left((B_0+C_0)\exp(i t)\right)_\mathrm{rms}}{1/\sqrt{2}},
\end{equation}
for longitudinal flow; and 
\begin{equation}
  b_\mathrm{eff\perp} = \frac{\int_{0}^{1}w(0, z)_\mathrm{rms}dz}{\int_{0}^{1}\frac{\partial w(\infty, z)}{\partial y}_\mathrm{rms}dz} = \frac{\Re\left({\sqrt{\frac{iL^2\omega}{\nu}}(E_0-F_0)\exp(i t)}\right)_\mathrm{rms}}{1/\sqrt{2}}.
\end{equation}
for transverse flow \citep{Luchini1991, Maali,cottin2004}, where $\int_{0}^1\frac{\partial u_\mathrm{II}}{\partial y}(\infty,z) \mathrm d z = 1$. In viscous units, $b_0^+ = b_\mathrm{eff}Lu_\tau/\nu$, where $u_\tau$ is the friction velocity.
\begin{figure}
  \centering
  \begin{subfigure}{\textwidth}
    \includegraphics[scale=0.39]{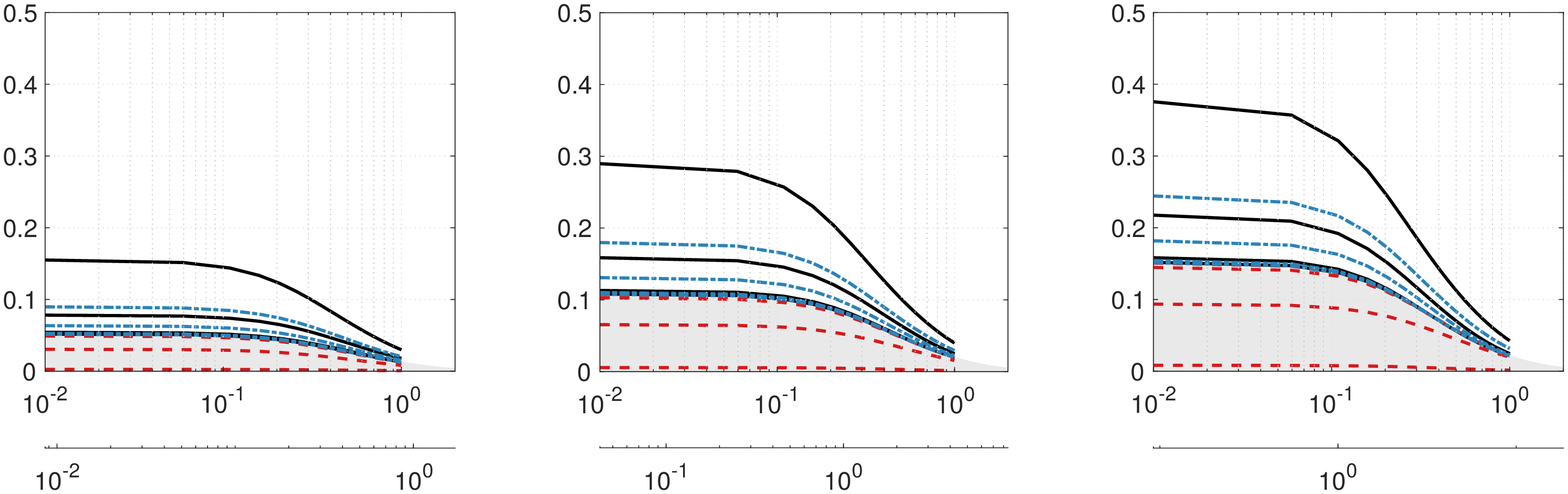}
    \put(-420,60){\rotatebox{90}{$b_\mathrm{eff\parallel}$}}
    \put(-350,-10){$L^2\omega/\nu$}
    \put(-220,-10){$L^2\omega/\nu$}
    \put(-90,-10){$L^2\omega/\nu$}
    \put(-350,12){$\omega\nu/u_\tau^2$}
    \put(-220,12){$\omega\nu/u_\tau^2$}
    \put(-90,12){$\omega\nu/u_\tau^2$}
    \put(-380,57){\vector(0,-1){10}}
    \put(-380,26){\vector(0,1){10}}
    \put(-250,70){\vector(0,-1){10}}
    \put(-250,30){\vector(0,1){10}}
    \put(-120,75){\vector(0,-1){10}}
    \put(-120,35){\vector(0,1){10}}
    \put(-310,103){(a)}
    \put(-180,103){(b)}
    \put(-48,103){(c)}
  \end{subfigure}
  \begin{subfigure}{\textwidth}
    \includegraphics[scale=0.39]{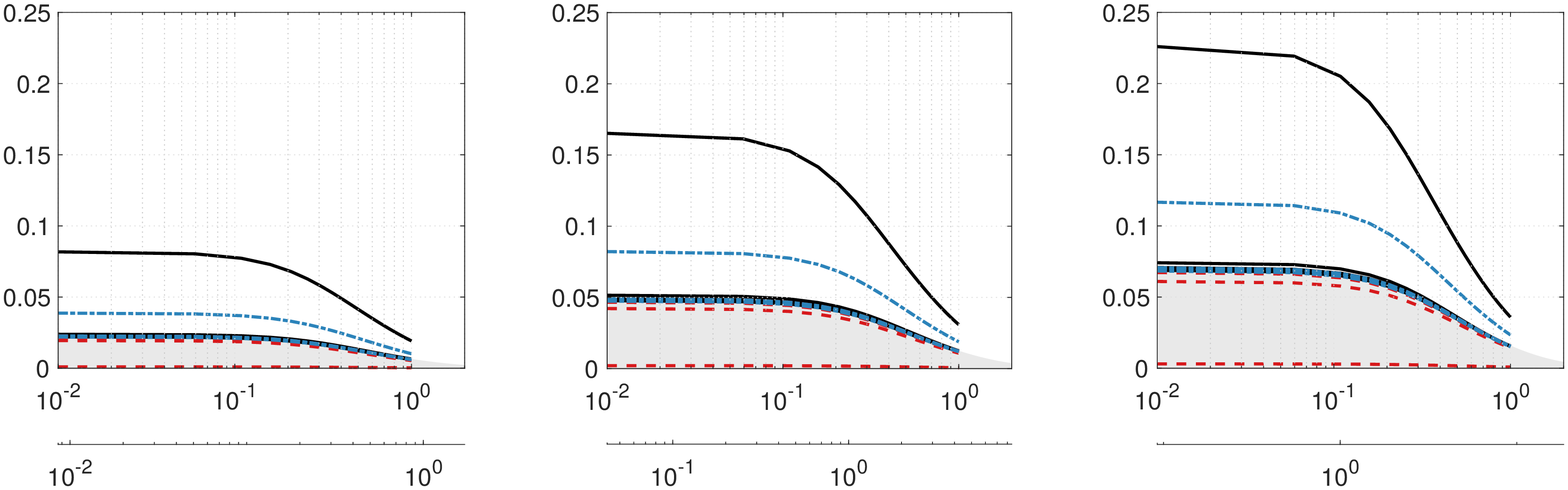}
    \put(-420,60){\rotatebox{90}{$b_\mathrm{eff\perp}$}}
    \put(-350,-10){$L^2\omega/\nu$}
    \put(-220,-10){$L^2\omega/\nu$}
    \put(-90,-10){$L^2\omega/\nu$}
    \put(-350,12){$\omega\nu/u_\tau^2$}
    \put(-220,12){$\omega\nu/u_\tau^2$}
    \put(-90,12){$\omega\nu/u_\tau^2$}
    \put(-380,57){\vector(0,-1){10}}
    \put(-380,26){\vector(0,1){10}}
    \put(-250,70){\vector(0,-1){10}}
    \put(-250,30){\vector(0,1){10}}
    \put(-120,75){\vector(0,-1){10}}
    \put(-120,35){\vector(0,1){10}}
    \put(-310,103){(d)}
    \put(-180,103){(e)}
    \put(-48,103){(f)}
  \end{subfigure}
\caption{Effective slip length $b_\mathrm{eff\parallel}$ (top row) and $b_\mathrm{eff\perp}$ (bottom row) with increasing frequencies of the forcing normalized by viscous units $u_\tau^2/\nu$ (top axis) or Womersley number $L^2\omega/\nu$ (bottom axis) with increasing representative (a) and (d) $Re_\tau=180$, (b) and (e) $Re_\tau=400$, (c) and (f) $Re_\tau=590$. Here, $\mu_r = 0.02,\ 0.37,\ 30.00$ are denoted by black solid lines, blue dashed-dotted lines, and red dashed lines respectively. The shaded area represents $b_\mathrm{eff}(L^2\omega/\nu, \mu_r\geq 1)$. The arrows indicate increasing penetration rate in the area of $\mu_r\geq 1$ or $\mu_r<1$ respectively.}
  \label{fig:beff}
\end{figure}
The unsteady effective slip length is plotted in figure~\ref{fig:beff}(a-c) for longitudinal flow and figure~\ref{fig:beff}(d-f) for transverse flow. Zero penetration of the external fluid with air pockets provide the highest slip length. Transverse effective slip length is more sensitive to the penetration of the external fluid: a $15\%$ penetration would result in an effective slip length overlapping with the fully-wetted grooved cases. 

\begin{figure}
  \centering
  \begin{subfigure}{\textwidth}
    \includegraphics[scale=0.39]{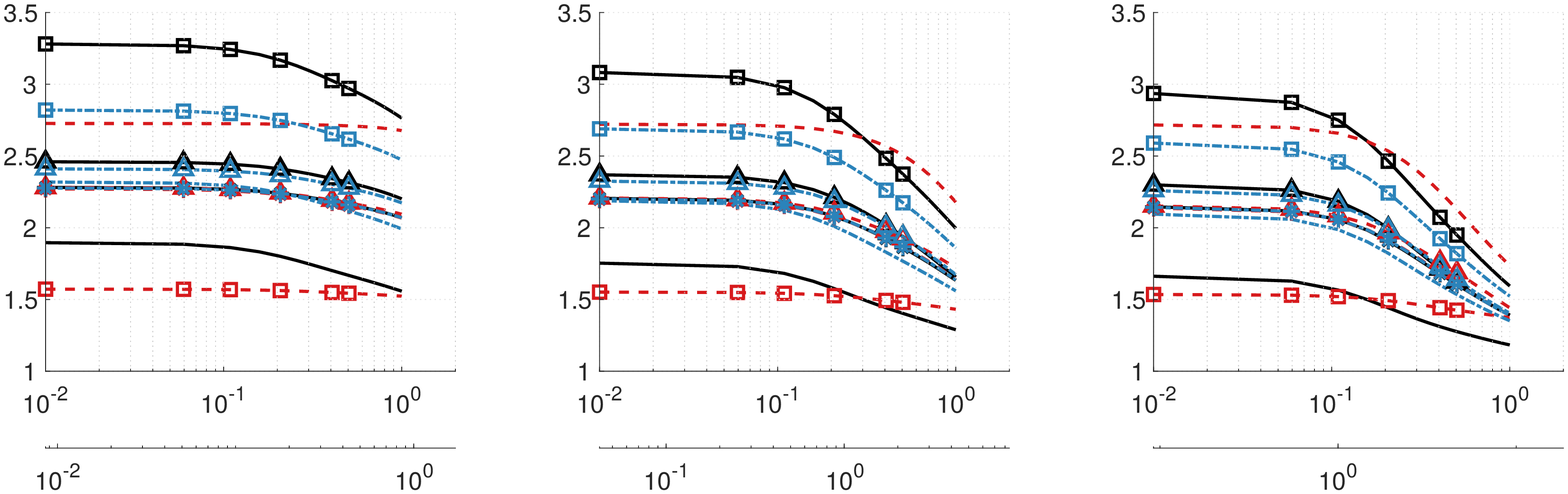}
    \put(-420,60){\rotatebox{90}{$b_\mathrm{eff\parallel}/b_\mathrm{eff\perp}$}}
    \put(-350,-10){$L^2\omega/\nu$}
    \put(-220,-10){$L^2\omega/\nu$}
    \put(-90,-10){$L^2\omega/\nu$}
    \put(-350,12){$\omega\nu/u_\tau^2$}
    \put(-220,12){$\omega\nu/u_\tau^2$}
    \put(-90,12){$\omega\nu/u_\tau^2$}
    \put(-310,103){(a)}
    \put(-180,103){(b)}
    \put(-48,103){(c)}
  \end{subfigure}
  \begin{subfigure}{\textwidth}
    \includegraphics[scale=0.39]{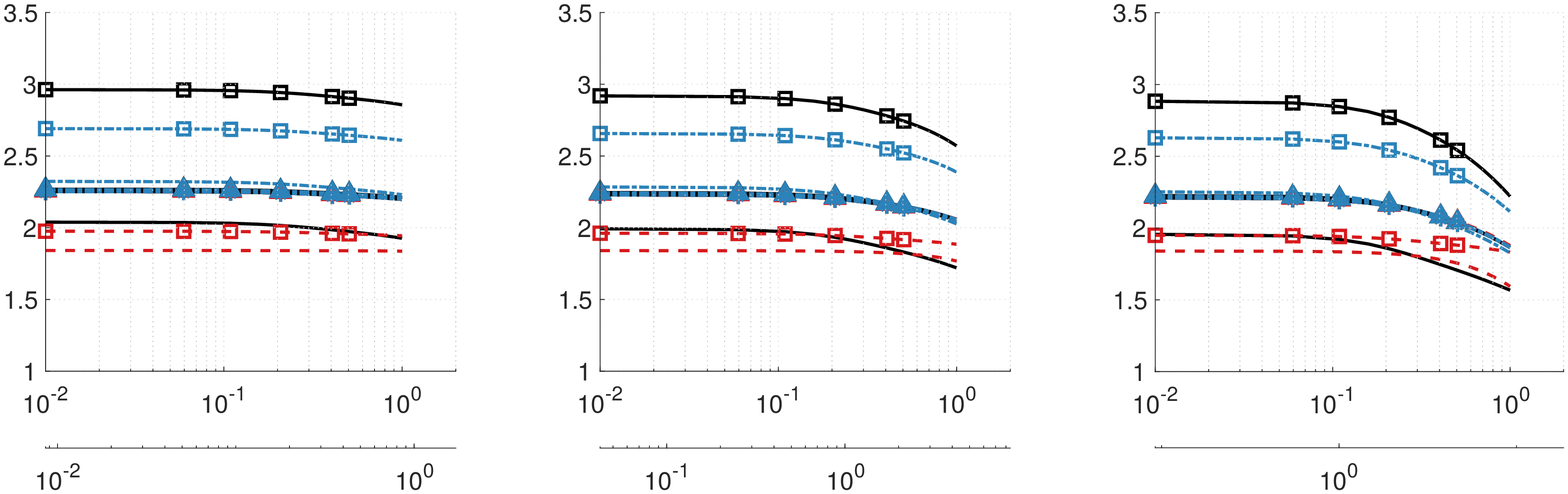}
    \put(-420,60){\rotatebox{90}{$b_\mathrm{eff\parallel}/b_\mathrm{eff\perp}$}}
    \put(-350,-10){$L^2\omega/\nu$}
    \put(-220,-10){$L^2\omega/\nu$}
    \put(-90,-10){$L^2\omega/\nu$}
    \put(-350,12){$\omega\nu/u_\tau^2$}
    \put(-220,12){$\omega\nu/u_\tau^2$}
    \put(-90,12){$\omega\nu/u_\tau^2$}
    \put(-310,103){(d)}
    \put(-180,103){(e)}
    \put(-48,103){(f)}
  \end{subfigure}
\caption{$b_\mathrm{eff\parallel}/b_\mathrm{eff\perp}$ with increasing frequencies of the forcing normalized by viscous units $u_\tau^2/\nu$ (top axis) or Womersley number $L^2\omega/\nu$ (bottom axis) with increasing representative (a) and (d) $Re_\tau=180$, (b) and (e) $Re_\tau=400$, (c) and (f) $Re_\tau=590$. Top row: $a = 0.875$; bottom row: $a = 0.5$. Here, $\mu_r = 0.02,\ 0.37,\ 30.00$ are denoted by black solid lines, blue dashed-dotted lines, and red dashed lines respectively. Penetration rate of $0$, $0.15$, $0.5$, $0.9$ is denoted by lines, $\square$, $\triangle$, $*$ respectively. The symbols differentiate lines and do not reflect actual data points.}
  \label{fig:bpara_bperp}
\end{figure}

Comparing the longitudinal effective slip length to the transverse values in figure~\ref{fig:bpara_bperp} using the ratio $b_\mathrm{eff\parallel}/b_\mathrm{eff\perp}$, it is found that the effective slip length for longitudinal flows is more than $1.5$ times of that of transverse flows with the same coverage ratio. The ratio drops at high frequency. A $15\%$ penetration rate could cause more decrease of the slip length in transverse direction, resulting a higher ratio. This agrees with the prior observation. But $b_\mathrm{eff\parallel}/b_\mathrm{eff\perp}$ is not a monotonic function of the penetration rate. For the combination of zero penetration and air-water interface, $b_\mathrm{eff\parallel}/b_\mathrm{eff\perp} = 2$ at low frequencies. This agrees with the finding in \cite{lauga2003} where no-shear strips in steady pipe flow were considered.

\begin{figure}
  \centering
  \begin{subfigure}{\textwidth}
    \includegraphics[scale=0.39]{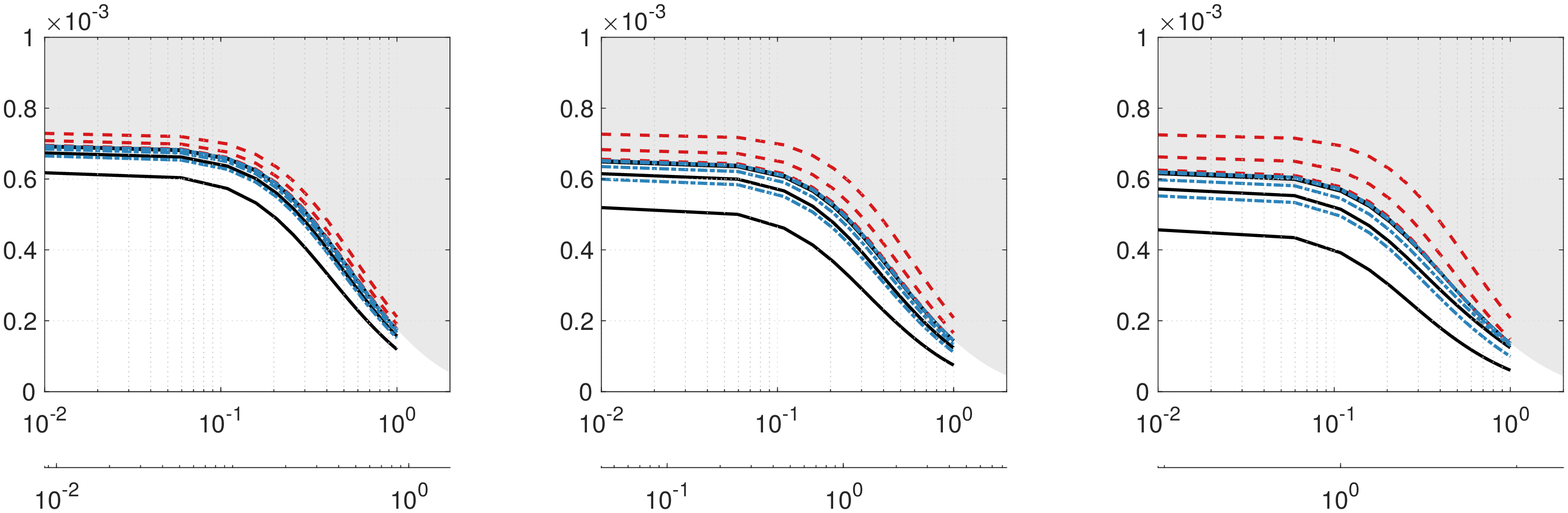}
    \put(-410,60){\rotatebox{90}{$\tau_{yx}$}}
    \put(-350,-10){$L^2\omega/\nu$}
    \put(-220,-10){$L^2\omega/\nu$}
    \put(-90,-10){$L^2\omega/\nu$}
    \put(-350,12){$\omega\nu/u_\tau^2$}
    \put(-220,12){$\omega\nu/u_\tau^2$}
    \put(-90,12){$\omega\nu/u_\tau^2$}
    \put(-380,90){\vector(0,1){10}}
    \put(-380,70){\vector(0,-1){10}}
    \put(-250,90){\vector(0,1){10}}
    \put(-250,70){\vector(0,-1){10}}
    \put(-120,90){\vector(0,1){10}}
    \put(-120,70){\vector(0,-1){10}}
    \put(-308,103){(a)}
    \put(-178,103){(b)}
    \put(-46,103){(c)}
  \end{subfigure}
  \begin{subfigure}{\textwidth}
    \includegraphics[scale=0.39]{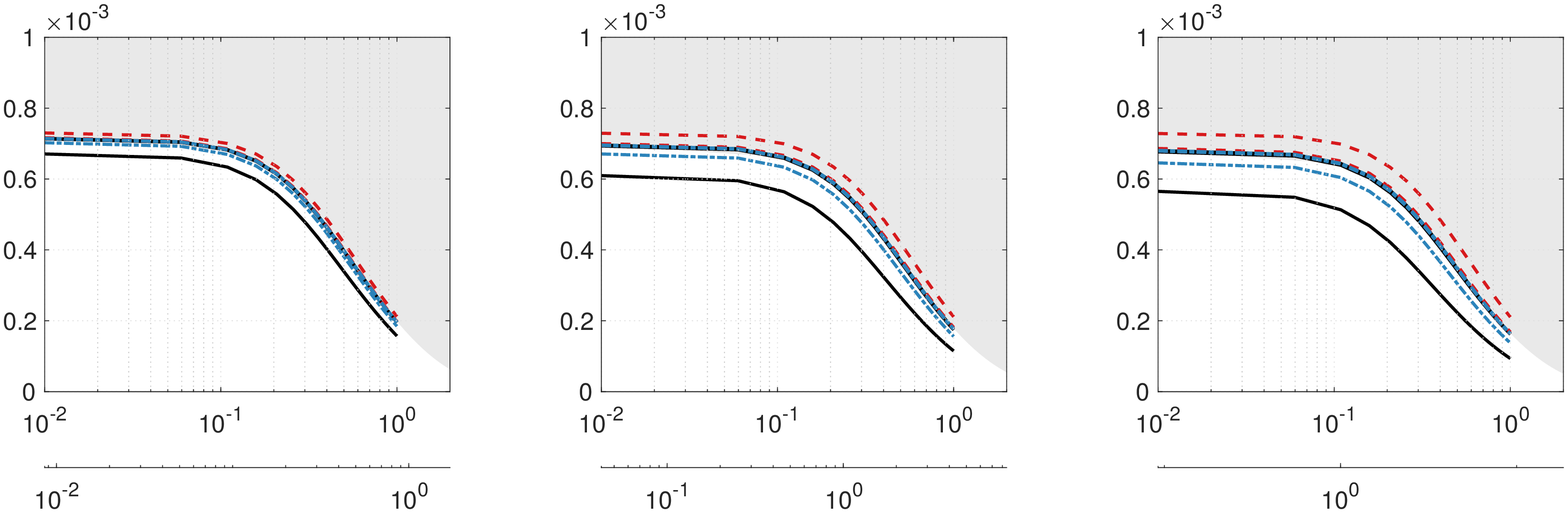}
    \put(-410,60){\rotatebox{90}{$\tau_{yz}$}}
    \put(-350,-10){$L^2\omega/\nu$}
    \put(-220,-10){$L^2\omega/\nu$}
    \put(-90,-10){$L^2\omega/\nu$}
    \put(-350,12){$\omega\nu/u_\tau^2$}
    \put(-220,12){$\omega\nu/u_\tau^2$}
    \put(-60,12){$\omega\nu/u_\tau^2$}
    \put(-380,90){\vector(0,1){10}}
    \put(-380,70){\vector(0,-1){10}}
    \put(-250,90){\vector(0,1){10}}
    \put(-250,70){\vector(0,-1){10}}
    \put(-120,90){\vector(0,1){10}}
    \put(-120,70){\vector(0,-1){10}}
    \put(-308,103){(d)}
    \put(-178,103){(e)}
    \put(-46,103){(f)}
  \end{subfigure}
  \caption{Shear stress $\tau_{yx}$ in longitudinal flow (top row) and $\tau_{yz}$ in transverse flow (bottom row) with increasing frequencies of the forcing normalized by viscous units $u_\tau^2/\nu$ (top axis) or Womersley number $L^2\omega/\nu$ (bottom axis) with increasing representative (a) and (d) $Re_\tau=180$, (b) and (e) $Re_\tau=400$, (c) and (f) $Re_\tau=590$. Here, $\mu_r = 0.02,\ 0.37,\ 30.00$ are denoted by black solid lines, blue dashed-dotted lines, and red dashed lines respectively. The shaded area represents $\tau(L^2\omega/\nu, \mu_r\geq 1)$. The arrows indicate increasing penetration rate in the area of $\mu_r\geq 1$ or $\mu_r<1$ respectively.}
  \label{fig:tau}
\end{figure}

The shear stress at the plane $y=0$ is evaluated by
\begin{equation}
  \tau_{yx} = \mu_1\Re\left(\sqrt{\frac{iL^2\omega}{\nu}}(B_0-C_0)\exp(i t)\right)_\mathrm{rms},
\end{equation}
for longitudinal flows, and
\begin{equation}
  \tau_{yz} = \mu_1\Re\left(\frac{iL^2\omega}{\nu}(E_0+F_0)\exp(i t)\right)_\mathrm{rms}.
\end{equation}
for transverse flows. Figure~\ref{fig:tau} shows the shear stress varying across the frequency range. Again, the combination of zero penetration and air pocket produces the least shear stress for both longitudinal and transverse flows. The effect of penetration of the external fluid has a larger impact on transverse flow than longitudinal flow. The trend observed in shear stress is consistent with those found in transfer function and effective slip length. As the frequency increases, the shear stress decreases, because the flow near the grooved surface is approaching a stationary state.

\section{Conclusion}
\label{sec:conclusion}
Analytical solutions of the multiphase Stokes flow generated by oscillatory velocity at a finite-height over a grooved surface have been derived. In the limit of the steady state, the analytical results agree well with the steady solution in \cite{schonecker2014} and \cite{Fu2017}. Numerical simulations using the VOF methodology were performed to validate the unsteady multiphase analytical solution. The analytical solution compares well with the DNS data. A transfer function based on the energy inside the grooved surface and the energy of the oscillatory input velocity was obtained. The analytical solution was parameterized by $\omega L^2/\nu$, a representative Reynolds number $Re_\tau$, and location of the multiphase interface $h/b$. We see that: (i) large values of $\omega L^2/\nu$ lower the transfer function, (ii) increasing $Re_\tau$ decreases the forcing height and therefore increases the transfer function, and (iii) When $\mu_r<1$, penetration of the external fluid suppresses the transfer function; when $\mu_r>1$, penetration of the external fluid increases it.

The variation of effective slip length and shear stress over the grooved plane show a consistent trend as the transfer function. The grooves show greater impedance to transverse flow than longitudinal flow, as is reported by other researchers \citep{Luchini1991, Choi1993}. Such an impedance is further strengthened when the external fluid penetrates into the groove. The ratio between the rms value of the longitudinal effective slip length $b_\mathrm{eff\parallel}$ and that of the transverse effective slip length $b_\mathrm{eff\perp}$ is optimal when $\mu_r = 0.02$, $1-h/b=0.15$, meaning that the surface is providing a large slip while impeding the cross-flow motion the least. When $\mu_r = 0.02$, $1-h/b=0$, $b_\mathrm{eff\parallel}/b_\mathrm{eff\perp}$ is $2$, which agrees with the finding of \cite{lauga2003}.

The dimensionless parameter $\omega L^2/\nu$ can be related to the $\Rey$ in a turbulent flow as: $\frac{\omega L^2}{\nu} = \omega^+Re_\tau^2\left(\frac{L}{\delta}\right)^2$, which is approximately the order of $O(10)$ for a moderate $\Rey$ considering $L/\delta\sim O(10^{-2})$. As $\Rey$ increases, at the same $\omega^+$, both streamwise and spanwise velocity fluctuation characterised as the oscillating velocity in the analytical solution, approach the surface, the energy of the fluctuation will influence the interface stronger in both directions.

\section*{Acknowledgement}
This work was supported by the United States Office of Naval Research (ONR) MURI (Multidisciplinary University Research Initiatives) program under Grant N00014-12-1-0874 managed by Dr. Ki-Han Kim. Computing resources were provided by the Minnesota Supercomputing Institute (MSI).
\bibliographystyle{jfm}
\bibliography{unsteady_stokes}

\end{document}